\documentclass[12pt,preprint]{aastex}
\usepackage{emulateapj5}

\newcommand{\HI}{H~{\sc i}} 
\newcommand{\HII}{H~{\sc ii}}
\newcommand{\kms}{${\rm km~s^{-1}}$}

\slugcomment{To appear in {\em The Astrophysical Journal}}

\shortauthors{MCCLURE-GRIFFITHS ET AL} 
\shorttitle{GALACTIC DISTRIBUTION OF HI SHELLS}

\begin{document} 

\title{The Galactic Distribution of Large \HI\ Shells}

\author{N.\ M.\ McClure-Griffiths\altaffilmark{1}} 
\affil{Australia Telescope National Facility, CSIRO, PO Box 76, Epping NSW
1710, Australia} 
\email{naomi.mcclure-griffiths@csiro.au}

\author{John M.\ Dickey}
\affil{Department of Astronomy, University of Minnesota, 116 Church St SE,
Minneapolis, MN 55455}
\email{john@astro.umn.edu}

\author{B.\ M.\ Gaensler} 
\affil{Harvard-Smithsonian Center for Astrophysics, 60 Garden Street MS-6,
Cambridge, MA 02138}
\email{bgaensler@cfa.harvard.edu}
\and
\author{A.\ J.\ Green}
\altaffiltext{1}{Bolton Fellow}
\affil{Astrophysics Department, School of Physics, Sydney University, NSW
2006, Australia}
\email{agreen@physics.usyd.edu.au}

\authoraddr{Address correspondence regarding this manuscript to: 
                N. M. McClure-Griffiths
                ATNF, CSIRO
                PO Box 76
		Epping NSW 1710
		Australia }
\begin{abstract}
We report the discovery of nineteen new \HI\ shells in the Southern Galactic
Plane Survey (SGPS).  These shells, which range in radius from 40~pc to
1~kpc, were found in the low resolution Parkes portion of the SGPS dataset,
covering Galactic longitudes $l=253\arcdeg$ to $l=358\arcdeg$. Here we give
the properties of individual shells, including positions, physical
dimensions, energetics, masses, and possible associations.  We also examine
the distribution of these shells in the Milky Way and find that several of
the shells are located between the spiral arms of the Galaxy.  We offer
possible explanations for this effect, in particular that the density
gradient away from spiral arms, combined with the many generations of
sequential star formation required to create large shells, could lead to a
preferential placement of shells on the trailing edges of spiral arms.
Spiral density wave theory is used in order to derive the magnitude of the
density gradient behind spiral arms.  We find that the density gradient away
from spiral arms is comparable to that out of the Galactic plane and
therefore suggest that this may lead to exaggerated shell expansion away
from spiral arms and into interarm regions.
\end{abstract}

\section{Introduction}
\label{sec:intro}
\HI\ shells, as a class of objects, largely determine the structure,
dynamics and evolution of the interstellar medium (ISM).  These massive
objects, which are usually detected as voids in the neutral hydrogen (\HI)
emission, range in size from tens to hundreds of parsecs, and in some cases
even kiloparsecs \citep[e.g.][]{rand96,deblok00}.  It is believed that \HI\
shells are formed through the combined effects of stellar winds and
supernovae, which input $10^{51}-10^{53}$ ergs of energy into the ISM,
ionizing the neutral medium and sweeping up a massive expanding shell.
Because of the significant energies involved, \HI\ shells may be the large,
deterministic structures needed to power the turbulent cascade of energy
seen in spatial power spectrum analyses of the Galaxy \citep{spangler01}.
In addition, with lifetimes on the order of tens of millions of years large
\HI\ shells long outlive the radiative lifetimes of their parent \HII\
regions and supernova remnants and hence can be used as fossils to study the
effects of star formation in the Galaxy.

In the nearby Large and Small Magellanic Clouds hundreds of \HI\ shells
dominate the structure of the \HI\ \citep{staveley-smith97,kim98}.  In the
Milky Way, however, the number of cataloged shells is considerably smaller.
Existing shell catalogs, for example the \citet{heiles79,heiles84} catalogs
focus on the Northern Galaxy and are limited by undersampled, low resolution
\HI\ surveys.  There are no comparable \HI\ shell catalogs for the Southern
Galaxy.  Additionally, though there have been many studies of individual
supershells \citep[e.g.][]{maciejewski96,heiles98}, there are only a few
studies of the global properties or distributions of Galactic \HI\ shells
\citep[e.g.][]{ehlerova96}.  Clearly Galactic \HI\ shell catalogs are in
need of updating and expansion to more global scales.  Recent \HI\ surveys,
such as the Southern and Canadian Galactic Plane Surveys \citep[SGPS and
CGPS;][]{mcgriff01a,taylor99}, are increasing the number of shells known in
the Milky Way.  These surveys, which image the \HI\ in large regions in the
Galaxy, benefit from coverage of spatial scales ranging from parsecs to
kiloparsecs, enabling more complete catalogs of Galactic \HI\ shells.  With
these surveys we hope to be able to determine the nature of many \HI\ shells
and the extent to which the ISM in the Milky Way is shaped by them.

Here we present nineteen new \HI\ shells found in the Southern Galactic
Plane Survey.  This catalog is limited to shells larger than one degree in
angular diameter.  The shells range in physical size from 40 pc to 700 pc
and are distributed throughout the Parkes SGPS region ($253\arcdeg \leq l
\leq 358\arcdeg$; $|b| \leq 10\deg$).  We focus here on the properties of
these shells and their distribution in the Galaxy.  In \S
\ref{subsec:criteria} we discuss the selection criteria for the current
catalog.  The observed and calculated properties of the shells are given in
\S \ref{subsec:props} and several individual shells are described briefly in
\S \ref{subsec:individ}. Selection biases in the catalog are discussed in \S
\ref{subsec:biases}. In \S \ref{sec:distribution} we focus on the Galactic
distribution of \HI\ shells, including those from the catalogs of
\citet{heiles79,heiles84}.  We explore shell properties in the context of
spiral structure, investigating the density structure of the \HI\ disk (\S
\ref{subsec:shells&spiral}).  Here we suggest that because of a density
gradient from spiral arms to interarm regions, shells on the edges of spiral
arms should attain larger sizes than shells in other regions.
\section{Observations and Analysis}
\label{sec:obs}
The data presented here are part of the Southern Galactic Plane Survey
\citep[SGPS;][]{mcgriff01a} and were obtained with the inner seven beams of
the Parkes multibeam system, a thirteen beam 21-cm receiver package at prime
focus on the Parkes 64 m Radiotelescope near Parkes NSW,
Australia\footnote{The Parkes Radio Telescope is part of the Australia
Telescope which is funded by the Commonwealth of Australia for operation as
a National Facility managed by CSIRO.}.  The data cover the region
$253\arcdeg \leq l \leq 328\arcdeg$, $|b| \leq 10\arcdeg$ and were obtained
by the process of mapping ``on-the-fly'', scanning through three degrees in
Galactic latitude while recording data in 5 s samples.  As described in
\citet{mcgriff01c}, the observations were made during four observing
sessions on 1998 December 15-16, 1999 June 18-21, 1999 September 18-27, and
2000 March 10-15.  The narrow line IAU standard calibration regions, S6 and
S9, were observed daily for bandpass and absolute brightness temperature
calibration.

To allow for robust bandpass calibration, the data were recorded in
frequency switching mode, switching between a center frequency of 1419.0 and
1422.125 MHz every 5s with a bandwidth of 8 MHz across 2048 channels.  For a
complete description of the bandpass calibration procedure the reader is
referred to McClure-Griffiths et al.\ (2000; in prep.)\nocite{mcgriff00}.
Absolute brightness temperature calibration was performed for each beam and
each polarization by multiplying the data by a calibration factor determined
from observations of the IAU standard regions, S6 and S9.  The data were
then shifted to the Local Standard of Rest (LSR) frame by applying a Doppler
correction in the form of a phase shift in the Fourier domain.  All
velocities given here are with respect to the LSR.

Finally, the calibrated data were imaged using {\em Gridzilla}, a gridding
tool created for use with the Parkes multibeam data and found in the
Australia Telescope National Facility (ATNF) subset of the AIPS++ package.
The gridding algorithm is described in detail in Barnes et al.\
(2001)\nocite{barnes01}.  The data were gridded using a weighted median
technique, assuming a beamwidth of 16\arcmin,  employing a Gaussian
smoothing kernel of FWHM 18\arcmin\ with a cutoff radius of 10\arcmin, and a
cellsize of 4\arcmin.  Off-line channels were used for continuum subtraction
in the image domain.  The data were imaged as ten cubes, each covering
approximately $16\arcdeg \times 20\arcdeg$ $(l\times b$).  The final
calibrated \HI\ cubes have an angular resolution of $\sim 16\arcmin$, a velocity
resolution of $0.82$~\kms, and a rms noise of $\sim 0.2$ K.

\section{The SGPS Large \HI\ Shell Catalog}
\label{sec:catalog}
The majority of known Galactic \HI\ shells were cataloged from surveys of
the Northern sky \citep[e.g.][]{heiles79,heiles84}.  \HI\ shells in the
Southern Galactic Plane, with its view of the inner Galaxy including the
Norma, Scutum-Crux, Sagittarius-Carina, and Perseus spiral arms, have not
been carefully studied.  Because we naively expect shells to be correlated
with star formation and star formation rates are highest in the inner Galaxy
and in the spiral arms, the Southern Galaxy should be replete with \HI\
shells.  The broad goal of this work is to provide a complete catalog of
\HI\ shells with which to investigate shell properties and evolution.  The
sample is needed before we can attempt to classify shell types, formation
mechanisms, and distributions.

\subsection{Selection Criteria}
\label{subsec:criteria}
New \HI\ shells were identified by eye in the Parkes \HI\ line cubes. The
current sample is limited to expanding shells of angular size one degree or
larger.  In order to be included in the catalog, \HI\ shell candidates must
meet three criteria.  Those criteria are:
\begin{itemize}
\item Shells can be first identified as approximately elliptical, well
defined voids in the \HI\ channel images.  The void must be observed over at
least four consecutive channels ($\Delta v \sim 3$~\kms) to be considered a
shell, rather than a filament.  These voids are also characterized by a
shell wall to shell interior brightness temperature contrast of a factor of
five or more.
\item Shell candidates must change in size with LSR velocity.  An ideal
spherical, expanding shell would appear as a series of rings in consecutive
\HI\ channel images.  These rings decrease in size away from the shell's
center, culminating in small disks - or caps - at the velocity extremes of
the shell.  Though none of the shells presented here are ideal, expanding
spheres, all shells change angular radius in consecutive channel images.
\item Velocity profiles through the apparent centers of shell candidates
must show a dip, flanked by peaks, as expected for a true shell. The dip is
the \HI\ void and the flanking peaks are the two caps of the expanding
shell.  The minimum of the velocity profile is identified with the center of
the shell, which is also usually the velocity at which the shell attains its
greatest angular size.
\end{itemize}

It is imperative that spectra demonstrating a shell-like bowl be compared
with spectra from other positions near the shell.  \HI\ spectra in the
Galaxy are complicated; in addition to shells, dips in the spectrum may be
due to absorption by cold gas or decreases in emission due to lower gas
densities in regions between spiral arms, among other things.
Distinguishing between interarm regions and \HI\ shells can largely be
accomplished by examining spectra taken adjacent to the shell.  If these
spectra do not show a similar dip in the profile then the \HI\ void is
likely due to a shell or absorption.  Distinguishing between absorption and
a shell is more difficult.  In both cases the dip in the profile is
localized to the region identified as a void in the channel images.
\citet{mintner01} note that a shell can be distinguished from
self-absorption by subtle details of the velocity profile.  Self-absorption
features typically have narrower widths with steeper walls. Random cloud
motions will blur and broaden the edges of an emission profile bowl
created by a shell.  Finally, the relatively small change in velocity
with distance throughout most of the Galaxy means that cold clouds of a
finite size will have narrow velocity widths.  By contrast, the velocity
width of a shell is determined by its expansion velocity and is therefore
much broader.

\subsection{Shell Properties}
\label{subsec:props}
\begin{figure}
\centering
\begin{center}
\plotone{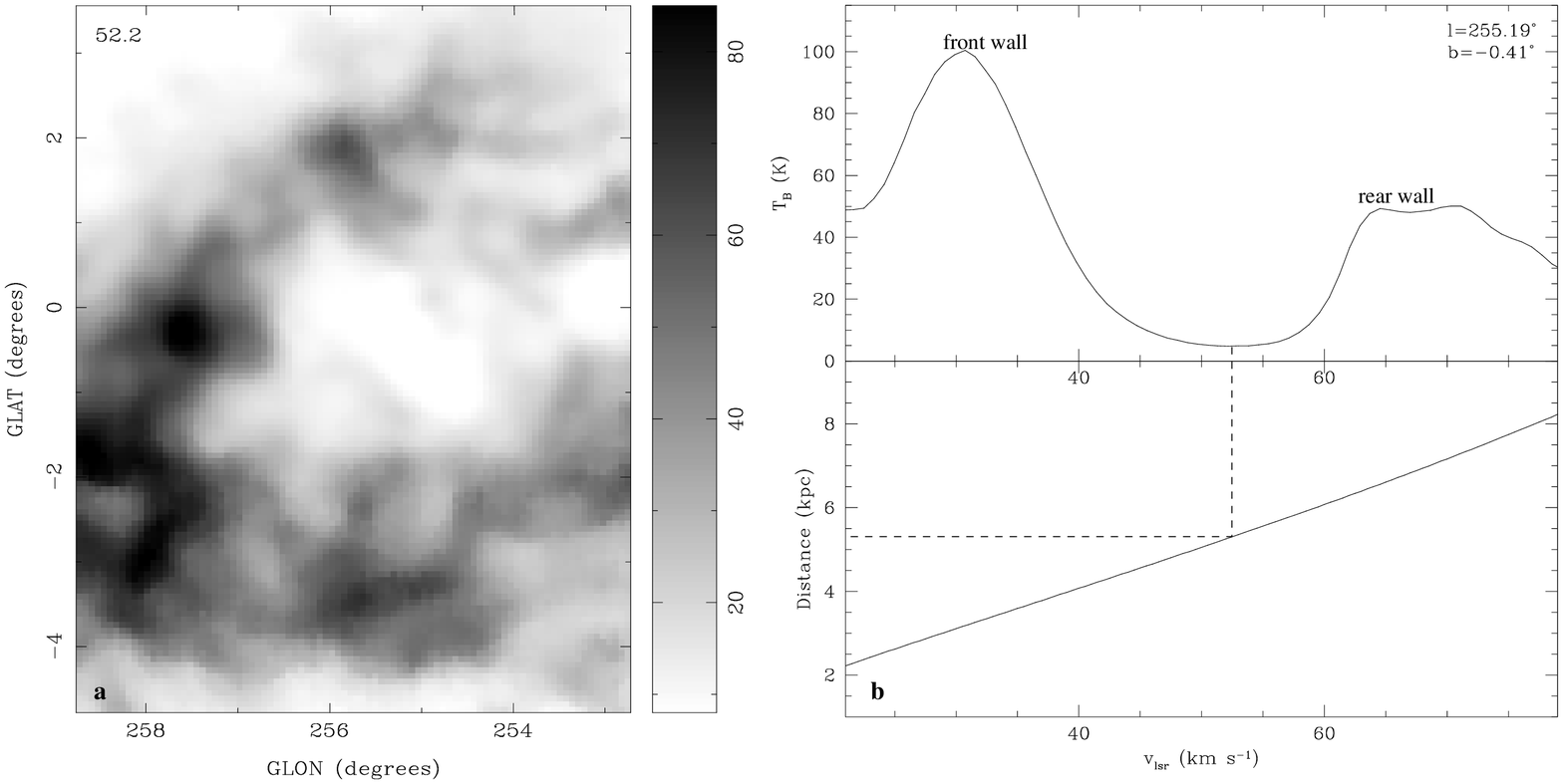}
\caption[GSH 255-00+52]{Grey-scale channel image of GSH 255-00+52 at
$v=51.3$ \kms\ ({\em a}) and velocity profile through the shell, with the
corresponding rotation curve along the line of sight ({\em b}).  The
grey-scale is linear in brightness temperature as shown on the wedge at the
right.  The central velocity of the shell is marked with dashed lines that
show the corresponding distance.
\label{fig:255-00+52}}
\end{center}
\end{figure}

\begin{figure}
\begin{center}
\plotone{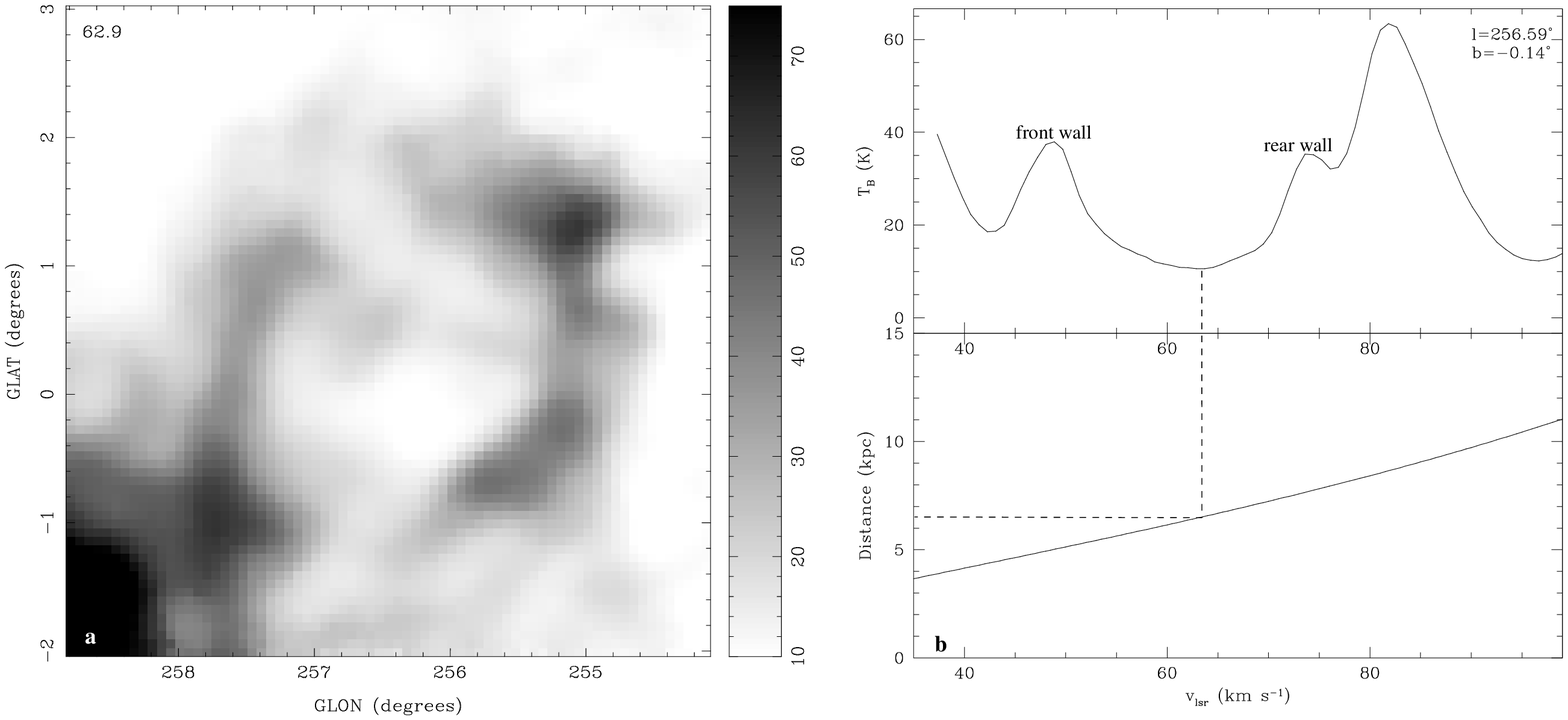}
\caption[GSH 256+00+63]{Grey-scale channel image of GSH 256+00+63 at
$v=62.9$ \kms\ ({\em a}) and velocity profile through the shell, with the
corresponding rotation curve along the line of sight ({\em b}).  The
grey-scale is linear in brightness temperature, as shown on the wedge at the
right.  The central velocity of the shell is marked with dashed lines that
show the corresponding distance.
\label{fig:256+00+63}}
\end{center}
\end{figure}

\begin{figure}
\begin{center}
\plotone{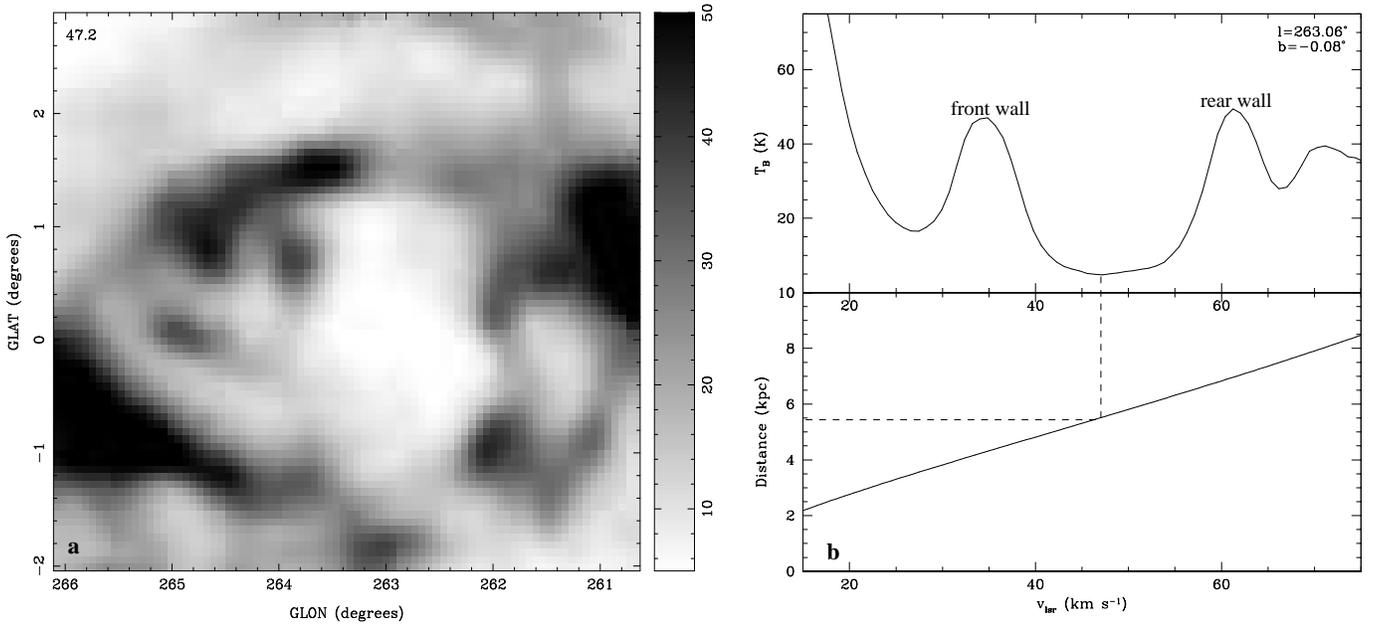}
\caption[GSH 263+00+47]{Grey-scale channel image of GSH 263+00+47 at
$v=47.2$ \kms\ ({\em a}) and velocity profile through the shell, with the
corresponding rotation curve along the line of sight ({\em b}).  The
grey-scale is linear in brightness temperature, as shown on the wedge at the
right.
\label{fig:263+00+47}}
\end{center}
\end{figure}

\begin{figure}
\begin{center}
\plotone{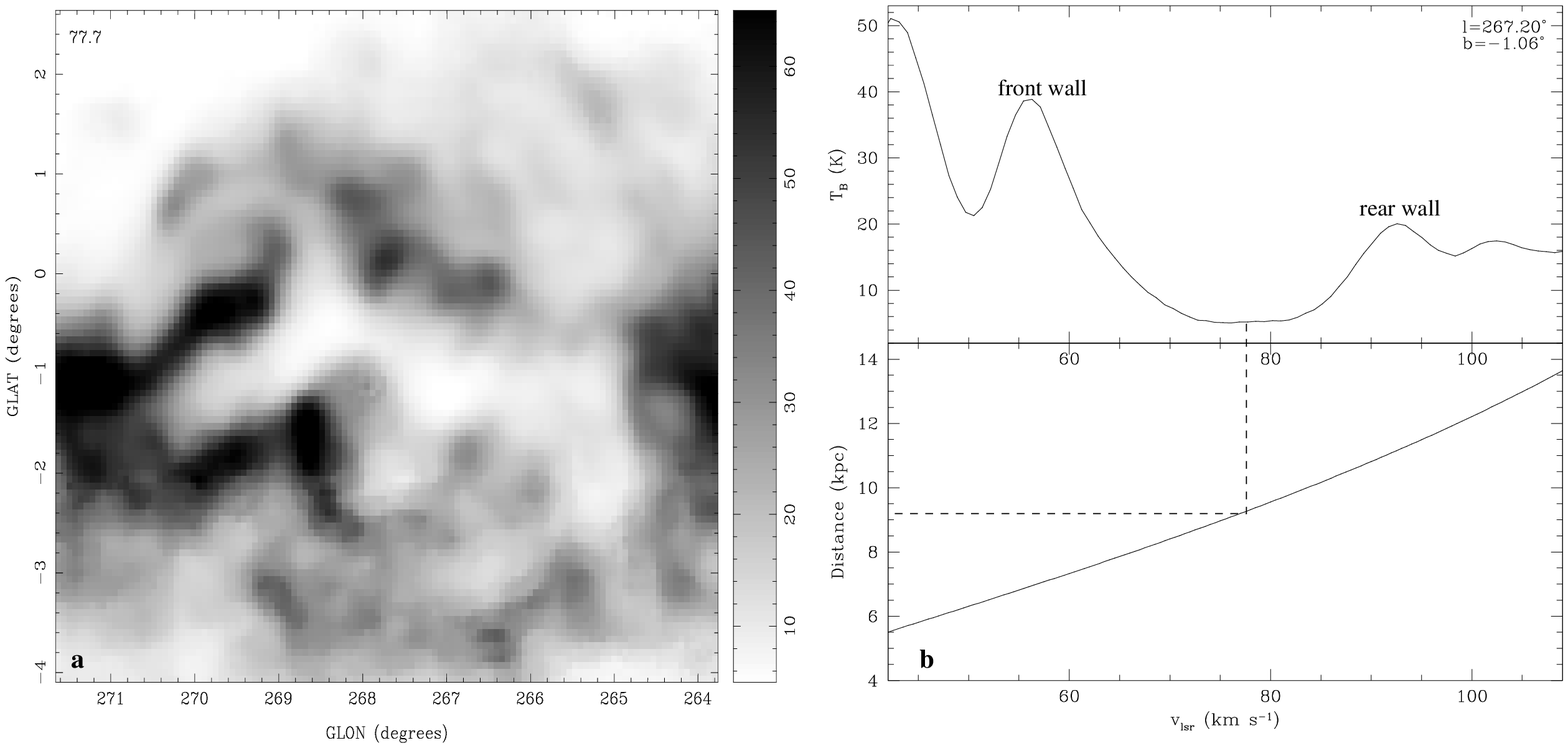}
\caption[GSH 267-01+77]{Grey-scale channel image of GSH 267-01+77 at
$v=77.7$ \kms\ ({\em a}) and velocity profile through the shell, with the
corresponding rotation curve along the line of sight ({\em b}).  The
grey-scale is linear in brightness temperature, as shown on the wedge at the
right.  The central velocity of the shell is marked with dashed lines that
show the corresponding distance.
\label{fig:267-01+77}}
\end{center}
\end{figure}

\begin{figure}
\begin{center}
\plotone{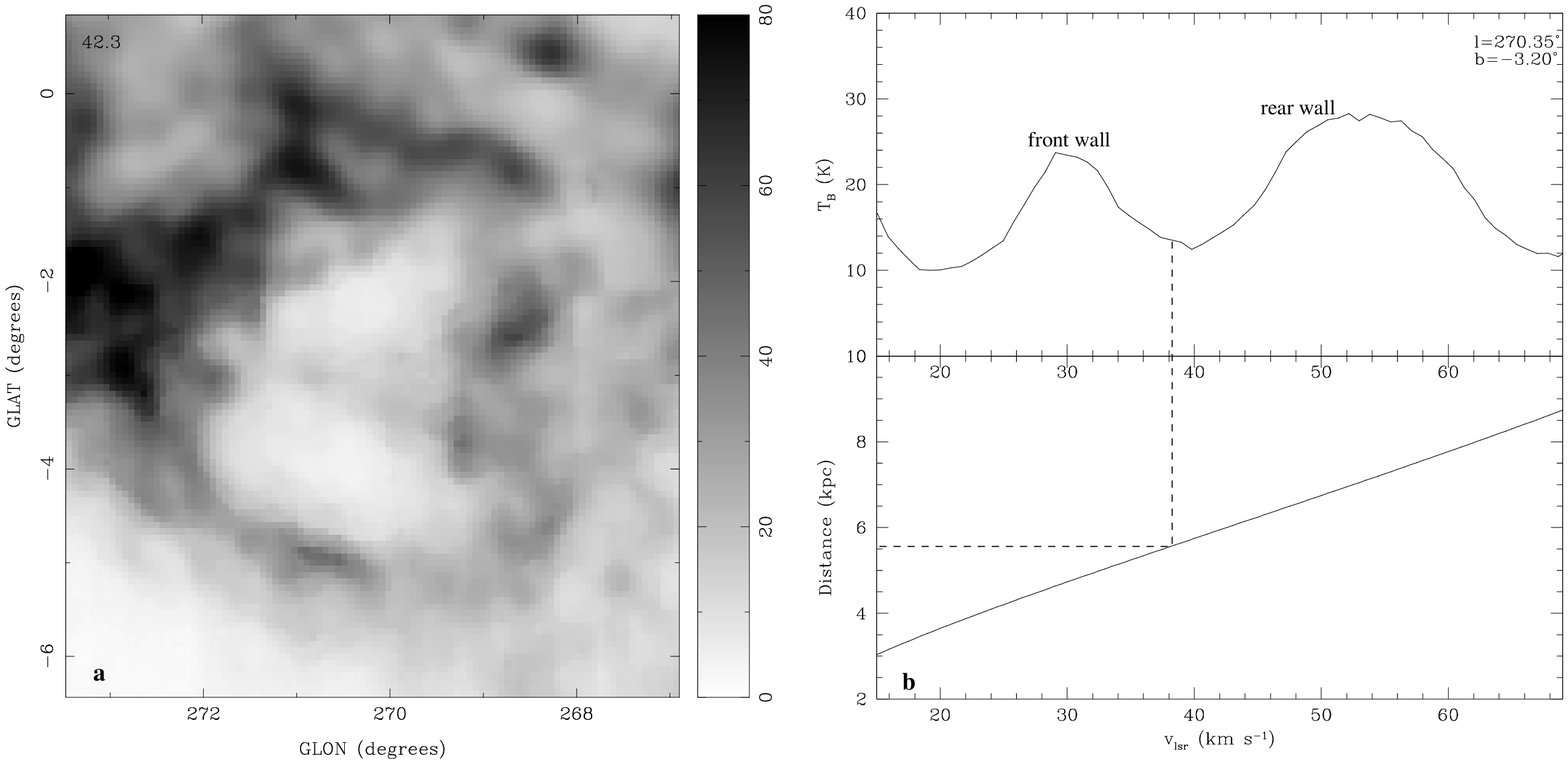}
\caption[GSH 270-03+42]{Grey-scale channel image of GSH 270-03+42 at
$v=47.3$ \kms\ ({\em a}) and velocity profile through the shell, with the
corresponding rotation curve along the line of sight ({\em b}).  The
grey-scale is linear as shown on the wedge at the right.  The central
velocity of the shell is marked with dashed lines that show the
corresponding distance.
\label{fig:270-03+42}}
\end{center}
\end{figure}

\begin{figure}
\begin{center}
\plotone{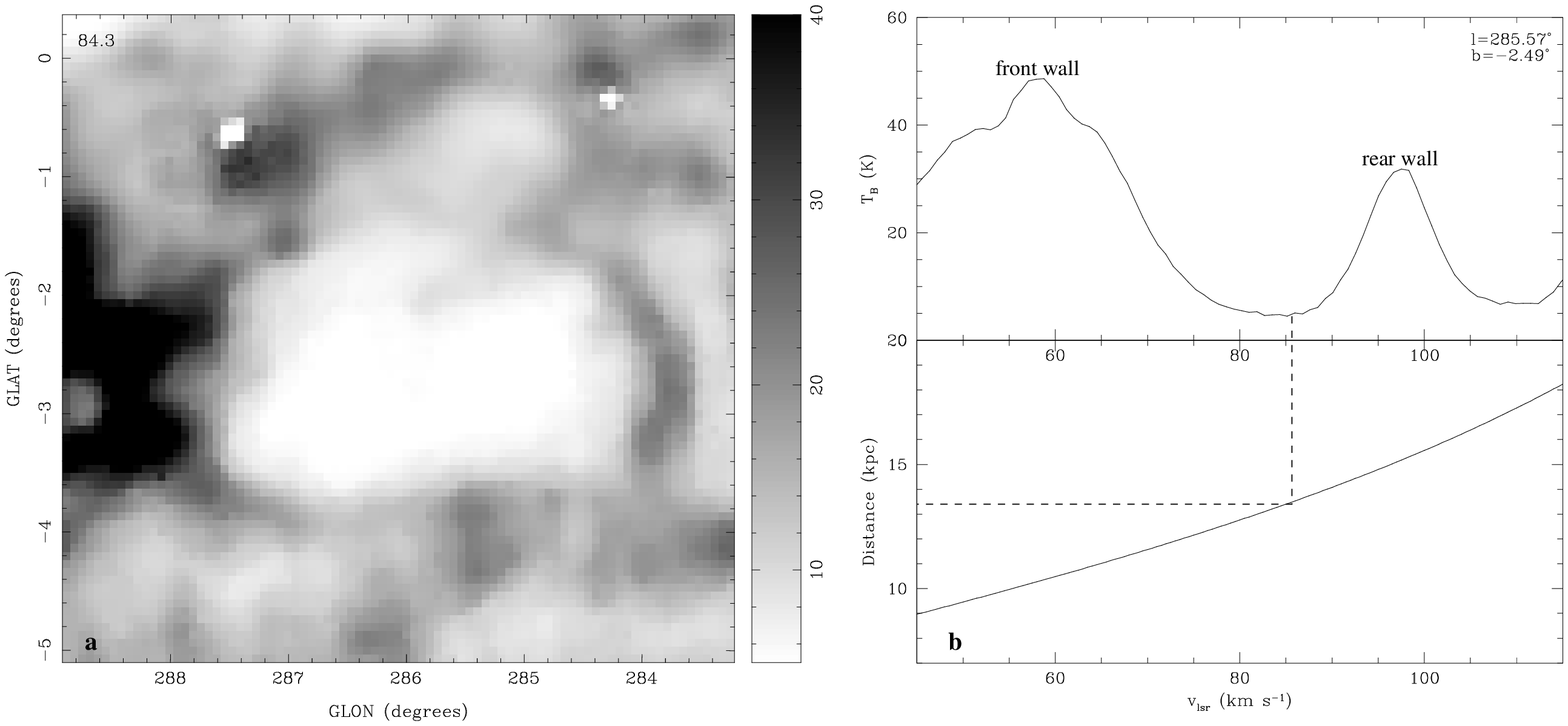}
\caption[GSH 285-03+86]{Grey-scale channel image of GSH 285-02+86 at
$v=86.0$ \kms\ ({\em a}) and velocity profile through the shell, with the
corresponding rotation curve along the line of sight ({\em b}).  The
grey-scale is linear in brightness temperature, as shown on the wedge at the
right.  The central velocity of the shell is marked with dashed lines that
show the corresponding distance.
\label{fig:285-02+86}}
\end{center}
\end{figure}

\begin{figure}
\begin{center}
\plotone{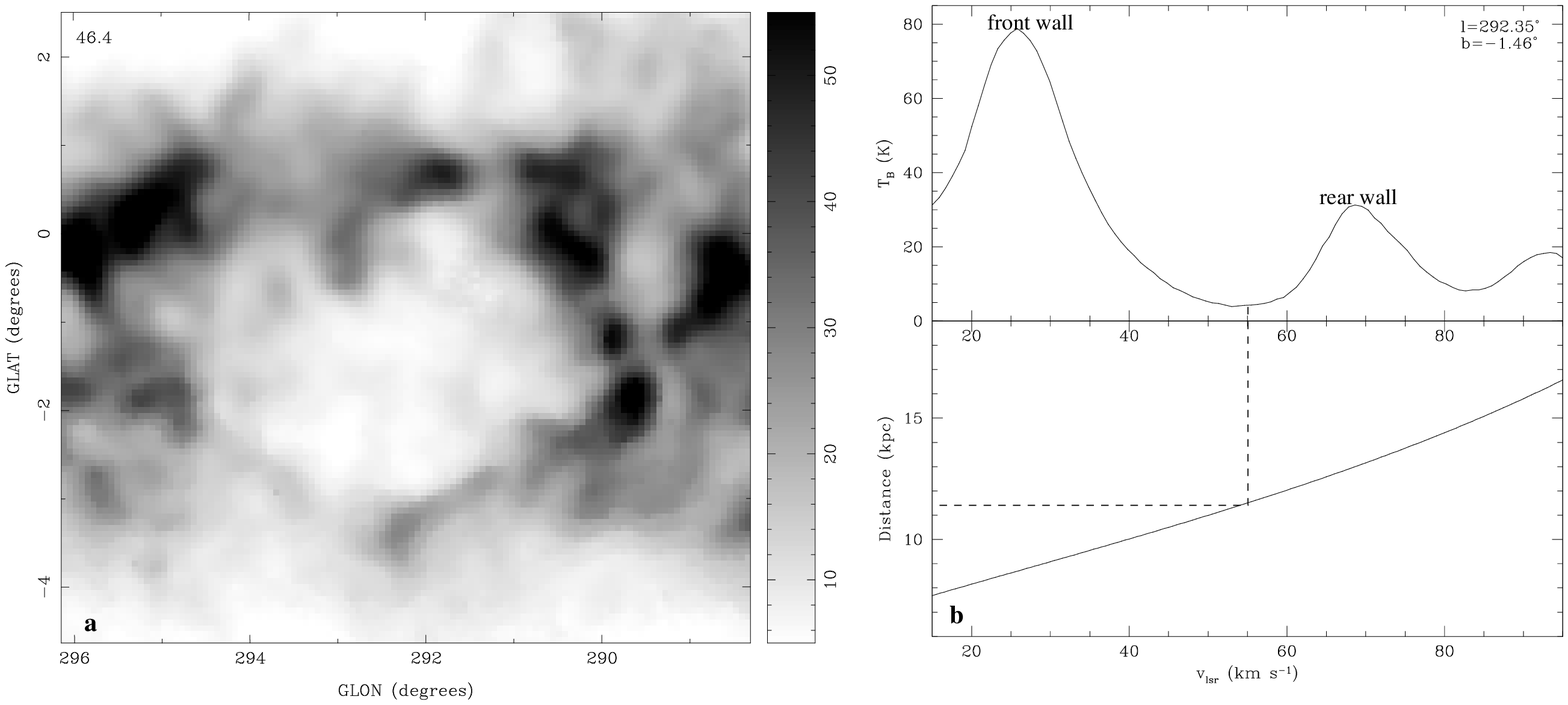}
\caption[GSH 292-01+46]{Grey-scale channel image of GSH 292-01+46 at
$v=46.4$ \kms\ ({\em a}) and velocity profile through the shell, with the
corresponding rotation curve along the line of sight ({\em b}).  The
grey-scale is linear in brightness temperature, as shown on the wedge at the
right.  The central velocity of the shell is marked with dashed lines that
show the corresponding distance.
\label{fig:292-01+46}}
\end{center}
\end{figure}

\begin{figure}
\begin{center}
\plotone{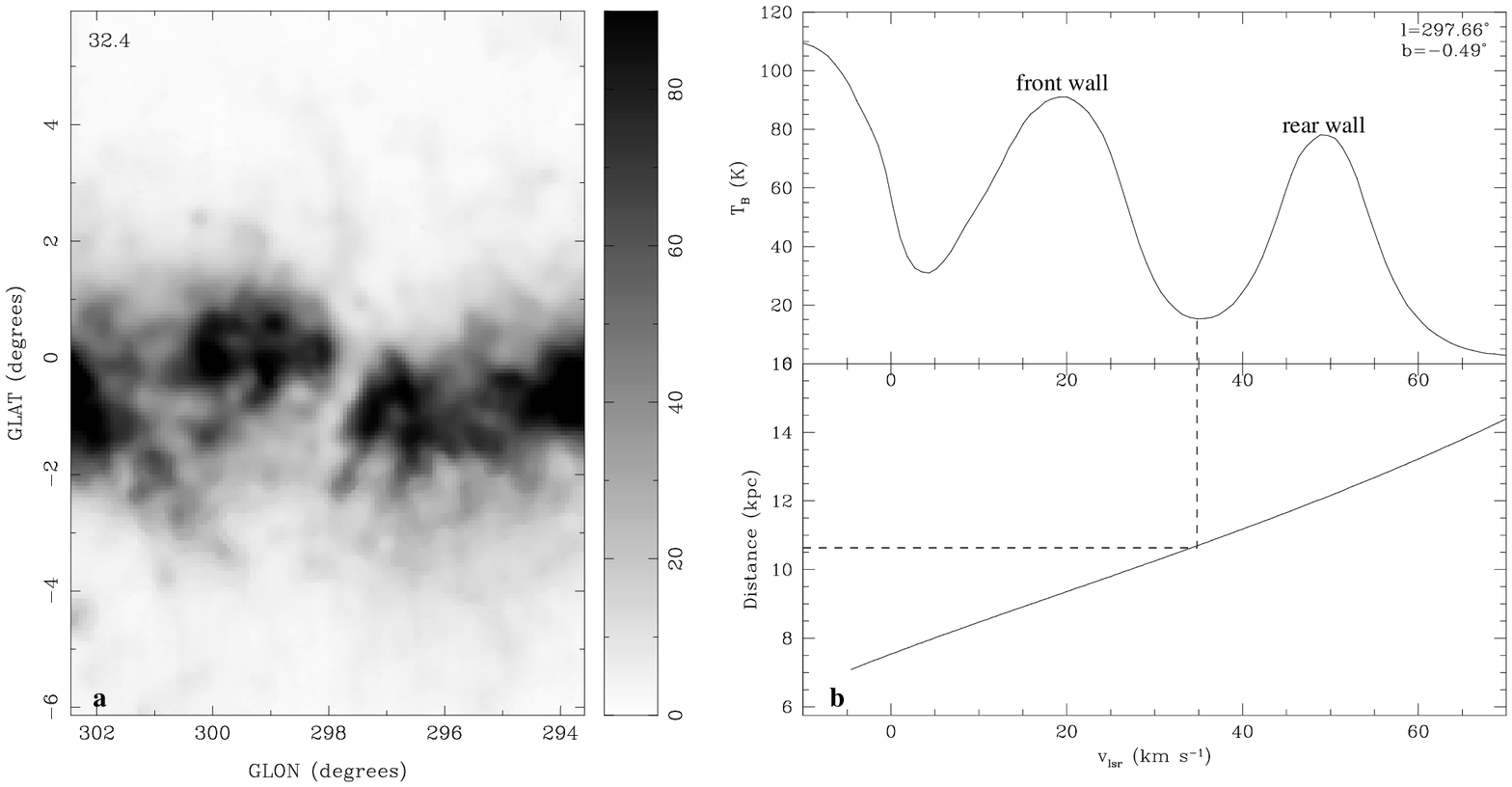}
\caption[GSH 298-01+35]{Grey-scale channel image of GSH 298-01+35 at
$v=32.4$ \kms\ ({\em a}) and velocity profile through the shell, with the
corresponding rotation curve along the line of sight ({\em b}).  The
grey-scale is linear in brightness temperature, as shown on the wedge at the
right.  The central velocity of the shell is marked with dashed lines that
show the corresponding distance.
\label{fig:298-01+35}}
\end{center}
\end{figure}

\begin{figure}
\begin{center}
\plotone{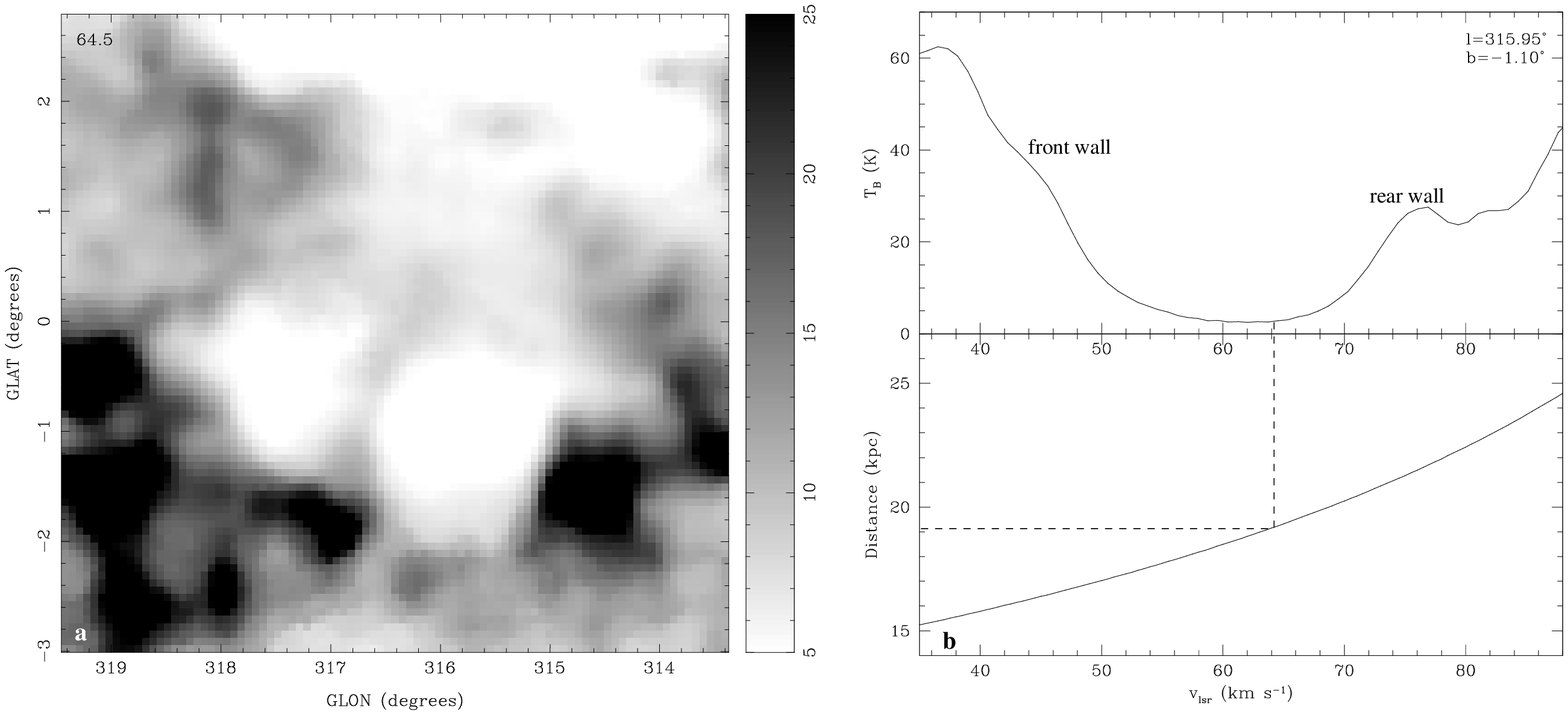}
\caption[GSH 316-00+65]{Grey-scale channel image of GSH 316-00+65 at
$v=64.5$ \kms\ ({\em a}) and velocity profile through the shell, with the
corresponding rotation curve along the line of sight ({\em b}).  The
grey-scale is linear in brightness temperature, as shown on the wedge at the
right.  The central velocity of the shell is marked with dashed lines that
show the corresponding distance.
\label{fig:316-00+65}}
\end{center}
\end{figure}

\begin{figure}
\begin{center}
\plotone{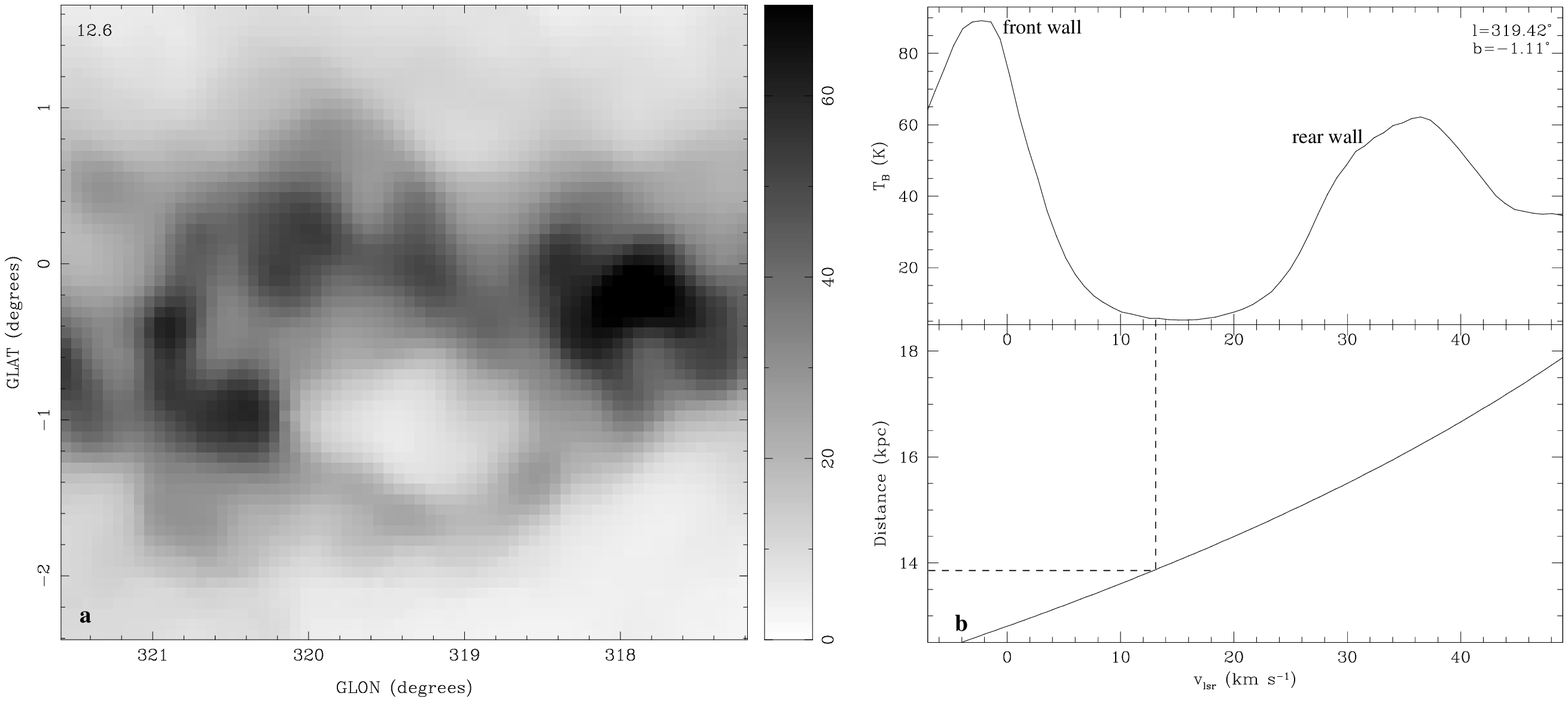}
\caption[GSH 319-01+14]{Grey-scale channel image of GSH 319-01+14 at
$v=12.6$ \kms\ ({\em a}) and velocity profile through the shell, with the
corresponding rotation curve along the line of sight ({\em b}).  The
grey-scale is linear in brightness temperature, as shown on the wedge at the
right. The central velocity of the shell is marked with dashed lines that
show the corresponding distance.
\label{fig:319-01+14}}
\end{center}
\end{figure}

\begin{figure}
\begin{center}
\plotone{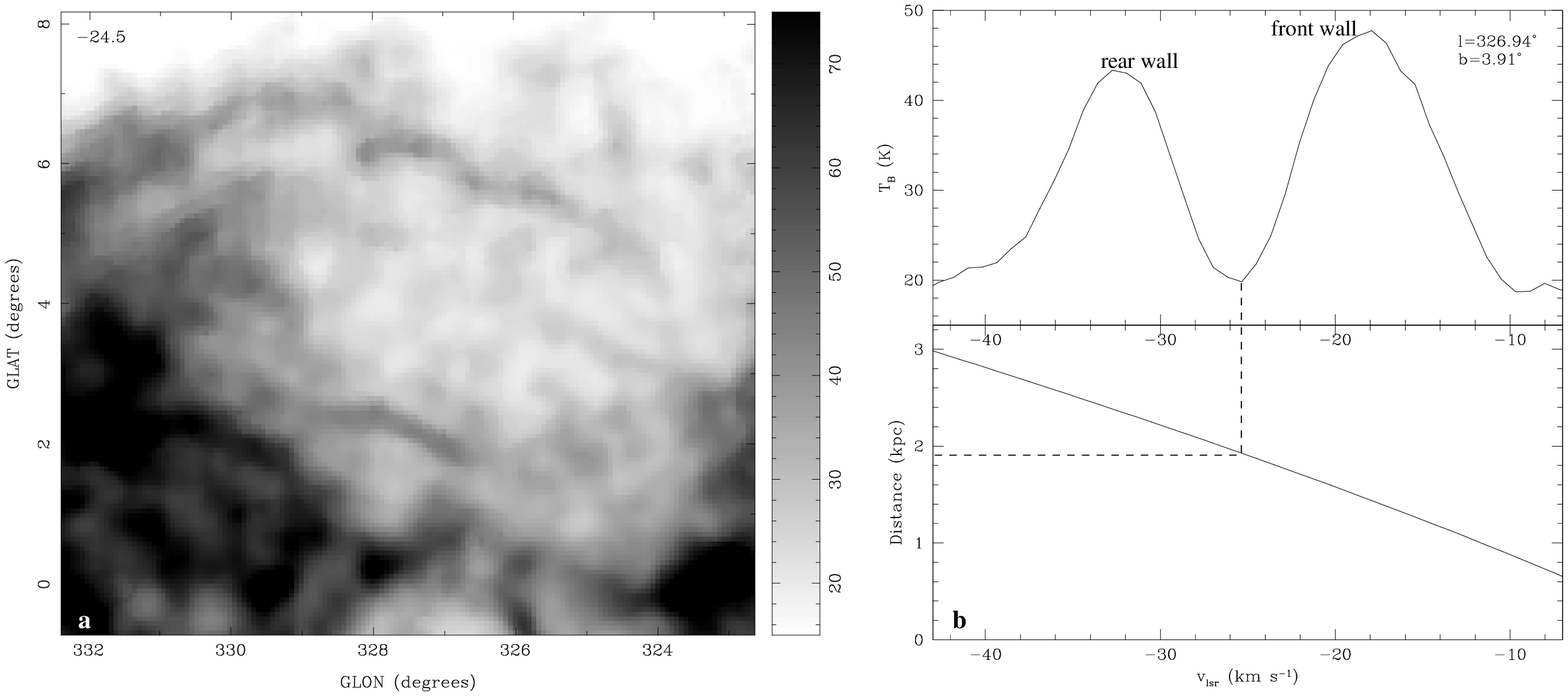}
\caption[GSH 327+04-25]{Grey-scale channel image of GSH 327+04-25 at
$v=-24.5$ \kms\ ({\em a}) and velocity profile through the shell, with the
corresponding rotation curve along the line of sight ({\em b}).  The
grey-scale is linear in brightness temperature, as shown on the wedge at the
right.  The central velocity of the shell is marked with dashed lines that
show the corresponding distance.
\label{fig:327+04-25}}
\end{center}
\end{figure}

\begin{figure}
\begin{center}
\plotone{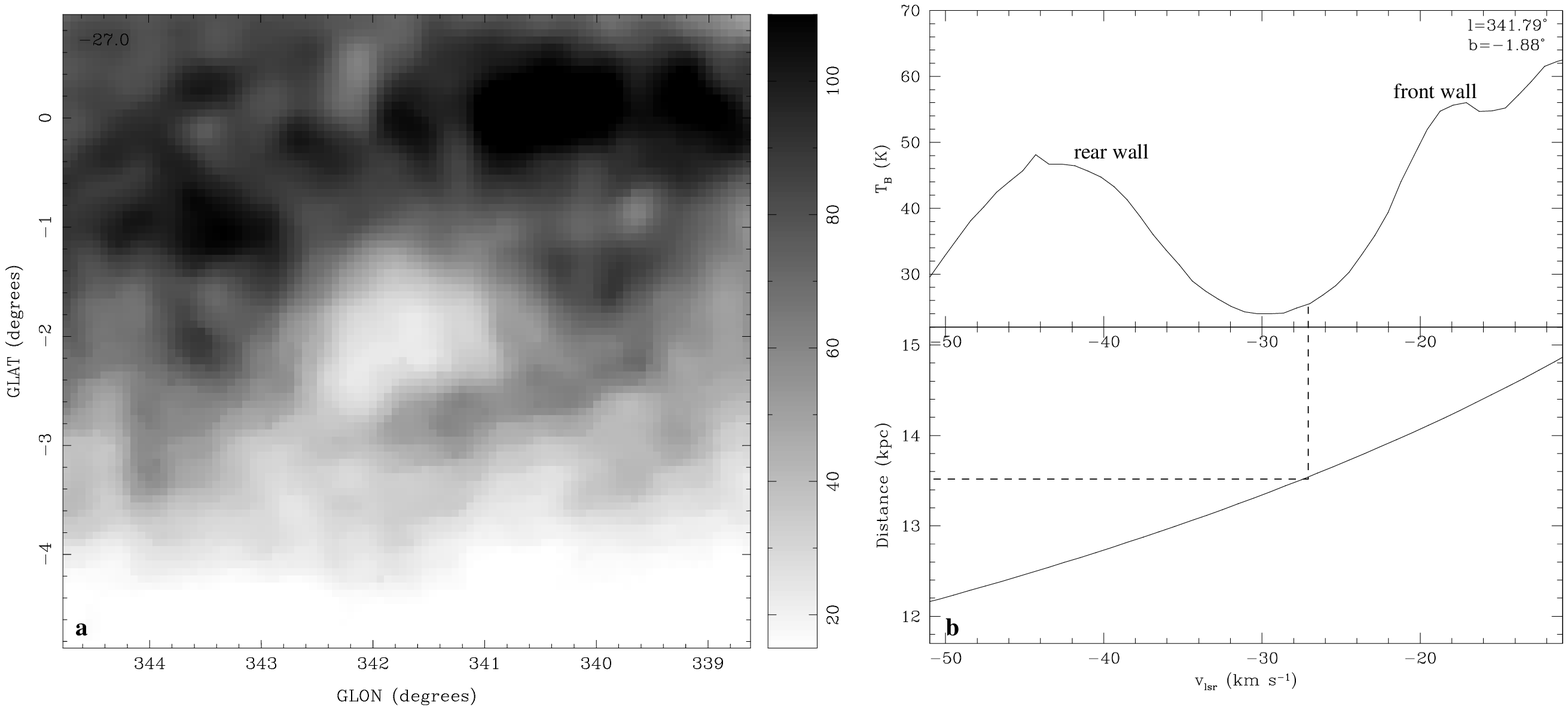}
\caption[GSH 342--02--27]{Grey-scale channel image of GSH 342--02--27 at
$v=-27.0$ \kms\ ({\em a}) and velocity profile through the shell, with the
corresponding rotation curve along the line of sight ({\em b}).  The
grey-scale is linear as shown on the wedge at the right.  The central
velocity of the shell is marked with dashed lines that show the
corresponding distance.
\label{fig:342-02-27}}
\end{center}
\end{figure}

\begin{figure}
\begin{center}
\plotone{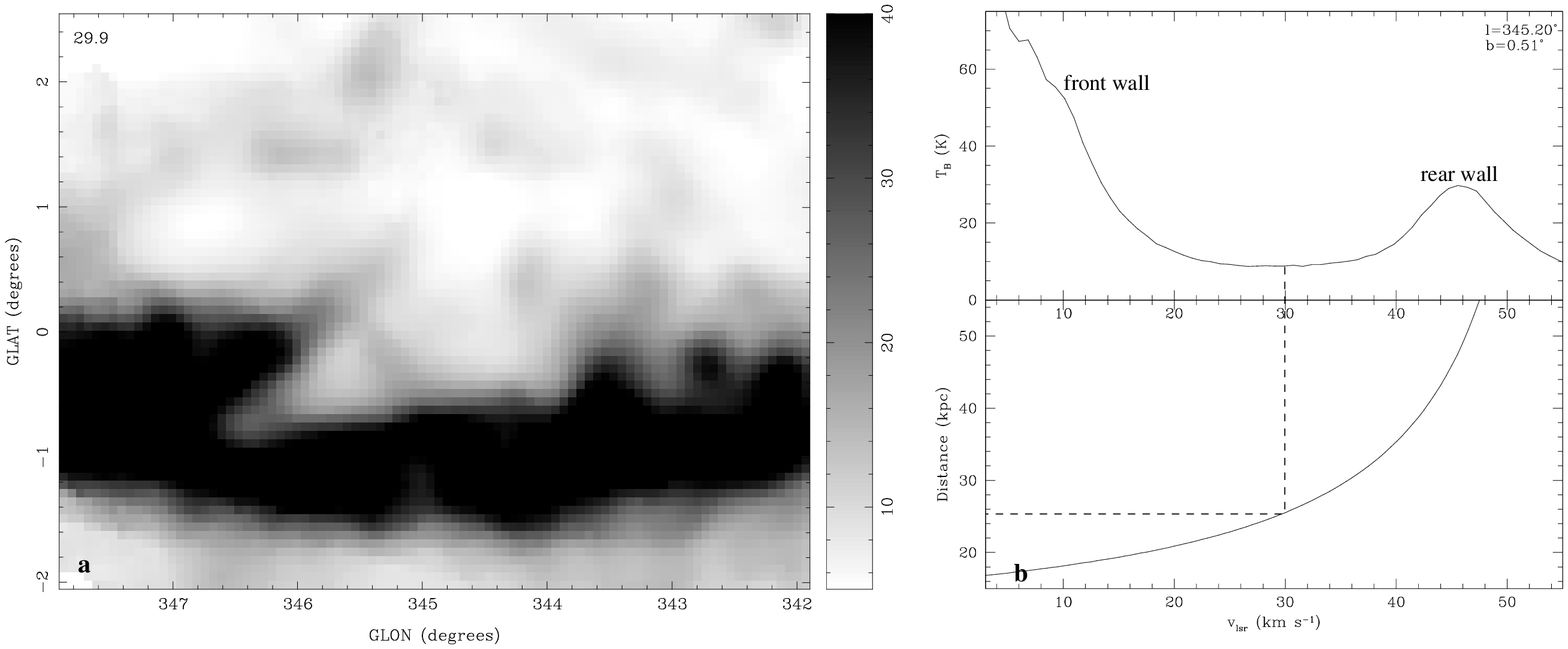}
\caption[GSH 345+00+30]{Grey-scale channel image of GSH 345+00+30 at
$v=29.9$ \kms\ ({\em a}) and velocity profile through the shell, with the
corresponding rotation curve along the line of sight ({\em b}).  The
grey-scale is linear in brightness temperature, as shown on the wedge at the
right.  The central velocity of the shell is marked with dashed lines that
show the corresponding distance.
\label{fig:345+00+30}}
\end{center}
\end{figure}

The SGPS shell catalog contains 19 new shells, four of which have been
described elsewhere \citep{mcgriff00,mcgriff01c}.  For each shell we list in
Table~\ref{tab:shells} the observed parameters: center position, angular
size, and expansion velocity.  We define the center of the shell in galactic
longitude, latitude, and velocity to be the point of minimum brightness
temperature nearest to the apparent geometric center.  Images of the shells
and their velocity profiles are shown in Figures \ref{fig:255-00+52} to
\ref{fig:345+00+30}.

The expansion velocity is estimated as half of the total velocity width of
the shell.  This is most easily determined from the velocity profile through
the center of the shell.  Using the peaks on either side of the shell bowl,
we determine a full velocity width for each shell.  Distinguishing the
fraction of a shell's velocity width that is due to expansion from that due
the spatial extent of the shell is difficult.  For each shell we calculate
the velocity gradient along the line of sight and compare the observed
velocity width of the shell to the inferred velocity width due to the
spatial extent of the shell.  In order to do this we assume that the shell
is nearly spherical and then calculate the expected velocity width from the
velocity gradient.  In all cases the velocity spread due only to the spatial
extent of the shell is less than 20\% of the observed shell velocity width.
We therefore assume that the observed width is due to expansion and estimate
that, to first order, the expansion velocity is half of the full velocity
width, $v_{\rm exp} \approx \Delta v /2$.

Kinematic distances and physical sizes are determined from the shell's
center velocity, using the rotation curve of \citet*{fich89}, and the
angular diameters of the major and minor axes of the shell.  We adopt the
IAU standard values for the Sun's orbital velocity, $\Theta_{\rm o} =
220$~\kms\ and Galactic center distance, $R_{\rm o}=8.5$~kpc.  Error
estimates for distance determinations assume that the dominant cause of
departures from Galactic circular rotation is streaming motions, which can
be as high as $\sim 10$~\kms\ \citep{burton88}.

The mass swept-up by the expanding shell is calculated from the column
density through the center of the shell, including all gas associated with
the feature.  We assume that the column density in the void is negligible
and calculate the mass from the mean column density near the center of the
shell.  The error in this calculation is based on the standard deviation of
the mean.  We assume that most of the mass in the near and far shell walls,
or ``caps,'' has been swept-up from the shell interior.  The column density
can then be used to estimate the ambient density of the medium into which
the shell expanded.  We assume a spherical geometry and use the axis in the
plane as the characteristic radius because the assumption of a constant
ambient density along the plane is more reasonable than perpendicular to the
plane, where the density will drop off with latitude.  Finally, the swept-up
mass is calculated from the shell volume and ambient gas density.  The
errors in column density and those in radius propagate through to the
ambient density and swept-up mass.

\HI\ shell energetics are often characterized by the expansion energy,
$E_{E}$, which is defined by \citet{heiles84} to be the equivalent energy
that would have to be instantaneously deposited at the shell center to
account for the shell's radius, $R_{\rm sh}$, and current rate of expansion,
$v_{\rm exp}$.  We give this energy as an alternative to the shell kinetic
energy, as derived by \citet{mccray87}, because it accounts for energy
losses due to radiative cooling.  Based on the \citet{chevalier74}
calculations for supernova remnant expansion, the expansion energy is given
by $E_{E} = 5.3 \times 10^{43} \, n_{\rm 0}^{1.12} \, R_{\rm sh}^{3.12} \,
v_{\rm exp}^{1.4}$~ergs, where $R_{sh}$ is in units of pc, $v_{exp}$ is in
\kms, and $n_0$ is in ${\rm cm^{-3}}$ \citep{heiles84}. The expansion energy
is strongly dependent on the shell radius and for a large shell with a
non-zero expansion velocity, the required formation energies can be very
large, $\sim 10^{53}$~ergs.

The ages of shells are very difficult to determine in the absence of an
observed power source.  As a substitute, the dynamic age, an estimate of the
age based on dynamic properties, is often quoted for shells.  One commonly
used technique is to calculate the dynamic age from the \citet{weaver77}
analytic solutions for a thin, expanding shell with a continuous rate of
energy injection.  An alternative, which we have adopted here, is to
calculate the dynamic age from the equations used to describe the evolution
of a supernova remnant in the late radiative phase.  In the radiative phase
the shock radius evolves with time as $R \propto t^{0.3}$
\citep*{chevalier74,cioffi88}.  So the dynamic age, $t_6$ in units of Myr,
for a shell of radius, $R_{sh}$ given in units of pc, and expansion
velocity, $v_{\rm exp}$ in units of \kms, is $t_6= 0.29 R_{sh} / v_{\rm
exp}$.  For comparison, the ages given here are a factor of $\sim 2$ less
than those calculated with the \citet{weaver77} equations.  The derived
parameters: physical dimensions, expansion energy, swept-up mass, and column
density are given in Table~\ref{tab:dershells}.
\subsection{Selected Individual Shells}
\label{subsec:individ}
For some shells, the properties were either exceptional or unusually complex
to calculate.  Here we describe those shells, explaining how the distances
and other properties were determined and in one case suggesting a possible
power source.  Images and detailed descriptions of GSH 277+00+36 and GSH
280+00+59, and GSH 304-00-12 and GSH 305+01-24 are given in
\citet{mcgriff00} and \citet{mcgriff01c}, respectively.  The remaining
shells are best demonstrated by the images and velocity profiles shown in
Figures \ref{fig:255-00+52} to \ref{fig:345+00+30}.
\subsubsection{GSH 267-01+77}
\label{sec:gsh267}
As seen in Figure \ref{fig:267-01+77}, the morphology of GSH 267-01+77 is
irregular.  The shell changes shape dramatically with velocity, but
maintains a very strongly emitting rim around the \HI\ void.  It is likely
that this shell was formed by the merger of several spatially distinct
shells.  However, the velocity profile is consistent throughout the shell
and has clear walls.  The physical axes of the full structure are $840\times
560$ pc and the expansion velocity is 18~\kms, which leads to an expansion
energy of $1.8\times 10^{53}$ ergs.  It is perhaps more reasonable to
consider the structure as three individual shells with radii on the order of
100 pc and expansion energies on the order of $\mbox{few} \times 10^{51}$
ergs.  Because the expansion energy goes as $R_{sh}^{3.14}$, considering the
system as three shells each of radius $R_{sh} = R_{\rm tot}/3$, the
expansion energy for the system, $E_E =E_1 + E_2 + E_3$, is decreased by a
factor of $\sim 10$.  This seems to eliminate the need for populous OB
associations at large galactocentric radii (GSH 267-01+77 is at $R_G=12.9$
kpc) by reducing the required number of supernovae and/or massive stellar
winds from $\sim 180$ to $\sim 18$.
\subsubsection{GSH 297-00+73}
\label{sec:gsh297}
\begin{figure}
\begin{center}
\plotone{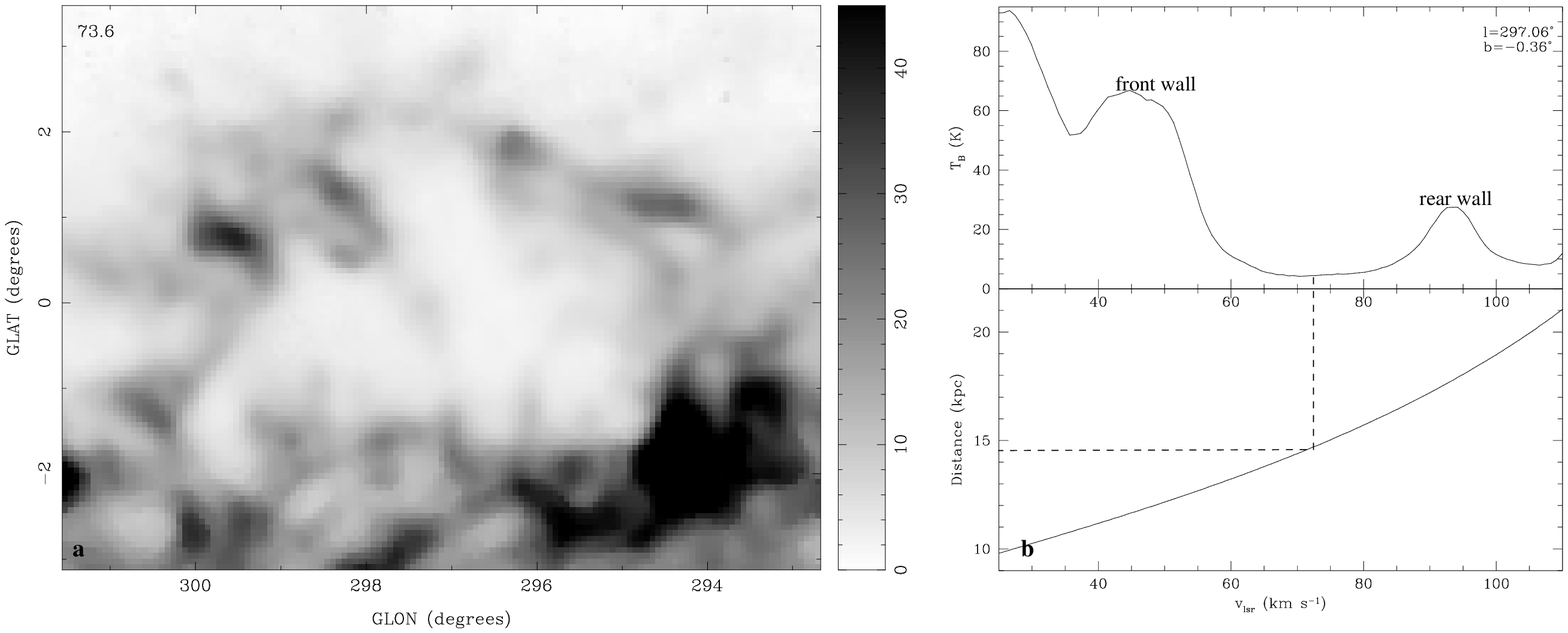}
\caption[GSH 297-00+73]{Grey-scale channel image of GSH 297-00+73 at
$v=73.6$ \kms\ ({\em a}) and velocity profile through the shell, with the
corresponding rotation curve along the line of sight ({\em b}).  The
grey-scale is linear in brightness temperature, as shown on the wedge at the
right. The central velocity of the shell is marked with dashed lines that
show the corresponding distance.
\label{fig:297-00+73}}
\end{center}
\end{figure}

GSH 297-00+73, shown in Figure \ref{fig:297-00+73}, also appears to be a
composite shell.  There is a narrow wall separating two original shells at
$l=297\fdg2$.  The wall is particularly strong near $v=63$~\kms\ and
$v=92$~\kms.  As seen in Figure~\ref{fig:297-00+73}, the wall disappears
almost completely near the center velocity of the composite structure.  The
decrease in emission may be due to an increased level of ionization from the
merger.  If we interpret the composite structure as two merged shells, one
shell of $R_1 =390$ pc, $v_{\rm exp1}=23$~\kms\ and another of $R_2 = 315$
pc, $v_{\rm exp2} = 20$~\kms, then the total expansion energy is $E_E=E_1
+E_2 = 1.9 \times 10^{53}$ ergs.  This shell is far away from any other
evidence of star formation, there are no \HII\ regions nearby nor any other
tracers of spiral arms.

\subsubsection{GSH 298-01+35}
\label{sec:gsh298}
Figure \ref{fig:298-01+35} shows GSH 298-01+35, a unique chimney or
``worm.''  The structure is very narrow in the longitudinal direction, but
extends several degrees out of the plane.  The chimney cuts through the
Galactic plane at a slight angle and then opens above and below the plane.
There is no evidence for closure within 6 degrees of the Galactic plane.
The kinematic distance is $10.5$ kpc, which places it near the edge of the
Sagittarius-Carina spiral arm. At this distance the chimney is only 75 pc
across but extends more than a kiloparsec above the plane.  The conditions
required to confine a shell so dramatically along the plane and still allow
for expansion into the halo are unclear.  The shell also has an unusual
velocity profile, in which the bowl is not very deep.

We have estimated the energy required to create a spherical cavity of radius
75 pc with an expansion velocity of 10~\kms\ as $\sim 1.1 \times 10^{52}$
ergs.  We add the caveat that this may be a severe underestimate if the
shell has been significantly confined along the plane. Near this position
and velocity, between $v=+16$~\kms\ and $v=+31$~\kms, are several \HII\
regions cataloged by \citet{caswell87}.  The closest \HII\ region,
G298.22-0.33, is at $v=+31$~\kms\ placing it near the center of the chimney.

\subsubsection{GSH 337+00--05}
\label{sec:gsh337}
\begin{figure}
\begin{center}
\plotone{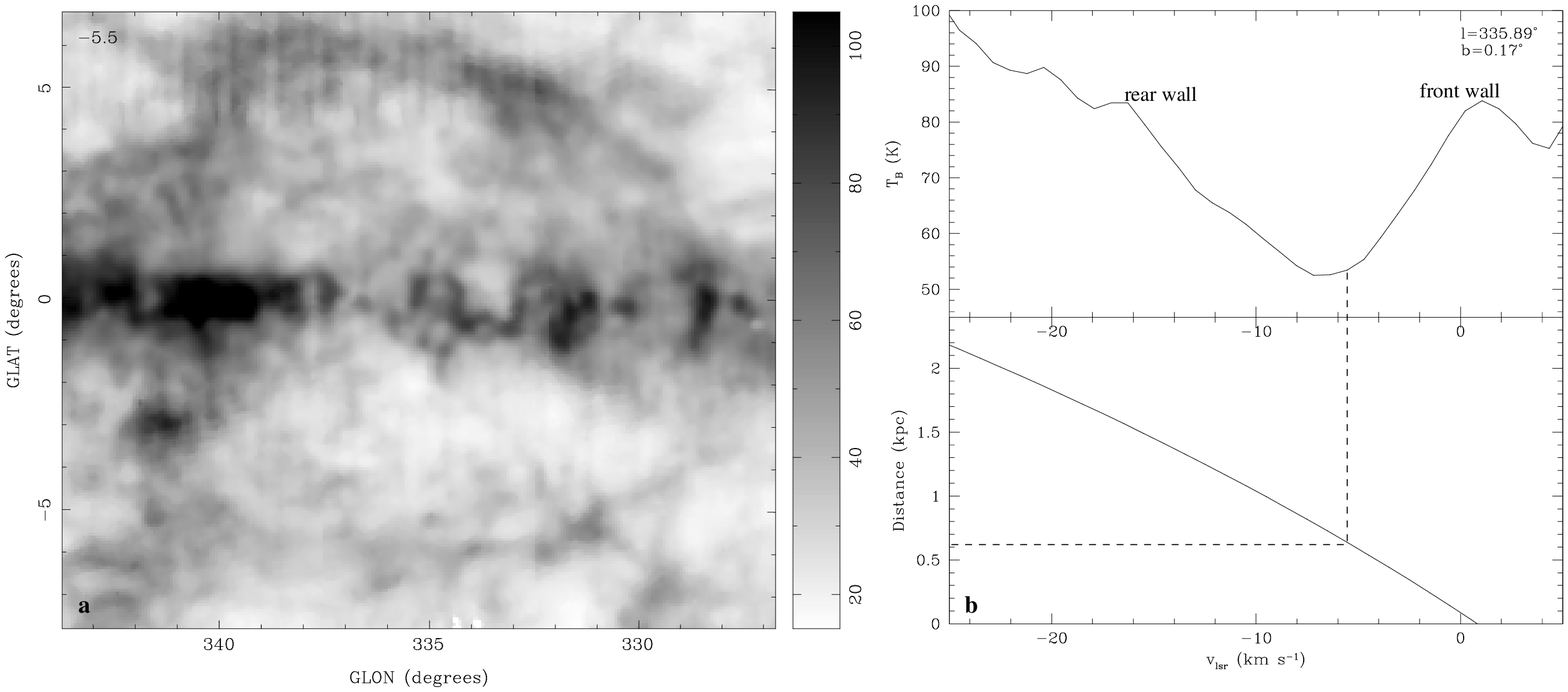}
\caption[GSH 337+00--05]{Grey-scale channel image of GSH 337+00--05 at
$v=-5.5$ \kms\ ({\em a}) and velocity profile through the shell, with the
corresponding rotation curve along the line of sight ({\em b}).  The
grey-scale is linear in brightness temperature, as shown on the wedge at the
right.  The central velocity of the shell is marked with dashed lines that
show the corresponding distance.
\label{fig:337+00-05}}
\end{center}
\end{figure}

Figure \ref{fig:337+00-05} is an image of the very large angular diameter
shell, GSH 337+00--05.  The shell has a very thin rim and regular elliptical
shape.  The \HI\ void is bisected by emission from the Galactic
Plane. Kinematic distances for GSH 337+00--05 are 570 pc or 15 kpc.  The
small radial velocity, however, makes these distances quite uncertain.  At
the near distance the shell has a size of $185 \times 125$ pc, whereas at
the far distance it would have a diameter of nearly 4 kpc.  Such a large
diameter is unlikely, therefore we place the shell at $\sim 570$ pc, on the
near edge of the Sagittarius-Carina arm.  Because of the shell's very local
systemic velocity, it is difficult to determine the expansion velocity.  The
local gas is largely filled, so the emission profile bowl (shown in
Fig. \ref{fig:337+00-05}) is not well-defined.  However, from the profile we
find an expansion velocity of 9~\kms\ with some uncertainty.  We assume an
ambient density of $1~{\rm cm^{-3}}$ and calculate a swept-up mass of
$1\times 10^5~{\rm M_{\sun}}$.  From the radius and expansion velocity an
expansion energy of $1.6 \times 10^{51}$ ergs is calculated.

The shell may be correlated with Ara OB1a, an association of 14 O, B, and A
stars at an adopted distance of 1.38 kpc.  The stars are distributed between
$l=335\arcdeg$ to $l=341\arcdeg$ and $b=\pm 3\arcdeg$, which places them in
the boundaries of shell.  Distance moduli for the individual stars indicate
distances between 790 and 2125 pc, with a mean distance of $1460\pm125$ pc.
Additionally, stars with measured radial velocities are between $-20$~\kms\
and $+5$~\kms, which agrees with the central velocity of the shell.  These
coincidences suggest that the cluster and shell are related.  If we assume
that the association has 14 stars each with a representative stellar wind
luminosity of $L_w \sim 6 \times 10^{35}~{\rm ergs~s^{-1}}$, corresponding
to that of a B0 star, we find that the expansion energy of the shell is
consistent with formation by this cluster with an age of $\sim 6$ Myr.
Because of large uncertainties in the kinematic distance of the \HI\ shell
are larger than the distance itself ($D=570\pm900$ pc) a definitive
identification is not possible.

\subsection{Selection Biases}
\label{subsec:biases}
Identifying shells by eye is a subjective process and results in a number of
biases in the compiled catalog.  The stringent selection criteria used in
this paper differ from those used for the \citet{heiles79,heiles84}
catalogs.  These criteria severely limit the number of shells included in
the catalog, but the confidence level of the classification is high.  There
remain many shell-like structures that do not meet all criteria and were
therefore not included in the catalog.  Because the structure of the \HI\ in
the outer Galaxy can be characterized as largely filamentary, it is
impossible to estimate the number of arched filaments that may be shells.
Hence this catalog significantly underestimates the number of shells,
perhaps by a factor or three or more.  Comparison with the surveys of the
Large and Small Magellanic Clouds \citep{kimphd,stanimirovic99}, for
example, would lead us to expect the identification of many more shells.

The number of \HI\ shells from this catalog found in the outer Galaxy
outnumbers those in the inner Galaxy by almost a factor of four.  This
occurs for a number of reasons.  Firstly, shells in the outer Galaxy are
considerably easier to detect than those in the inner Galaxy.  The distance
ambiguity results in a much higher filling factor for \HI\ emission at all
inner Galaxy velocities.  Objects that might otherwise appear as voids are
filled by emission from another Galactic position.  This selection bias is
exacerbated by the criterion that shells exhibit a characteristic bowl in
the velocity profile; candidate shells in the inner Galaxy often do not have
such a well defined dip in the velocity profile.  Secondly, the limited
latitude coverage of the survey restricts the number of nearby shells
detected. The cubes used to identify shells cover $16\arcdeg \times
22\arcdeg$ each.  Extremely large ($\gtrsim 10\arcdeg$) angular diameter
shells may extend beyond the edge of the cubes and are therefore difficult
to find.  The two extremely large angular diameter shells included in the
catalog, GSH 304-00-12 and GSH 337+00-05, are unusually continuous and
straightforward to detect.  Because longitudes towards the third quadrant
($l<310\arcdeg$) cover more of the outer Galaxy at comparatively nearby
distances, more shells can be detected.  Lines of sight at high longitudes
($l>340\arcdeg$) cross large distances interior to the solar circle before
reaching the outer Galaxy.  As a result, outer Galaxy shells at these
longitudes are very distant and therefore difficult to detect because of
their small angular sizes and weak emission.

It is difficult to overcome selection biases in \HI\ shell catalogs.  An
obvious solution would be to employ an automated searching algorithm.
Several attempts have been made to create automated searching routines for
\HI\ shell identification \citep[e.g.][]{thilker98}.  The \citet{thilker98}
technique is based on a three-dimensional cross correlation between a
simulated pattern and real data.  While this technique seems to be effective
in external galaxies, attempts to apply the technique to the Milky Way have
have not been very successful at detecting large shells.  The complicated
rotation curve, coupled with the fact that shells are typically
non-spherical and intertwined, makes it difficult to apply an automated
technique.  \citet{mashchenko99} have extended the three-dimensional
hydrodynamical simulations of shells to include more realistic models.  This
technique has been applied with moderate success in Canadian Galactic Plane
Survey (CGPS) data \citep{mashchenko99}.  However, the application has been
limited to shells around single stars for which the physics is more
straightforward than for supershells.

\section{Spatial Distribution of \HI\ Shells}
\label{sec:distribution}
\begin{figure}
\centering
\plotone{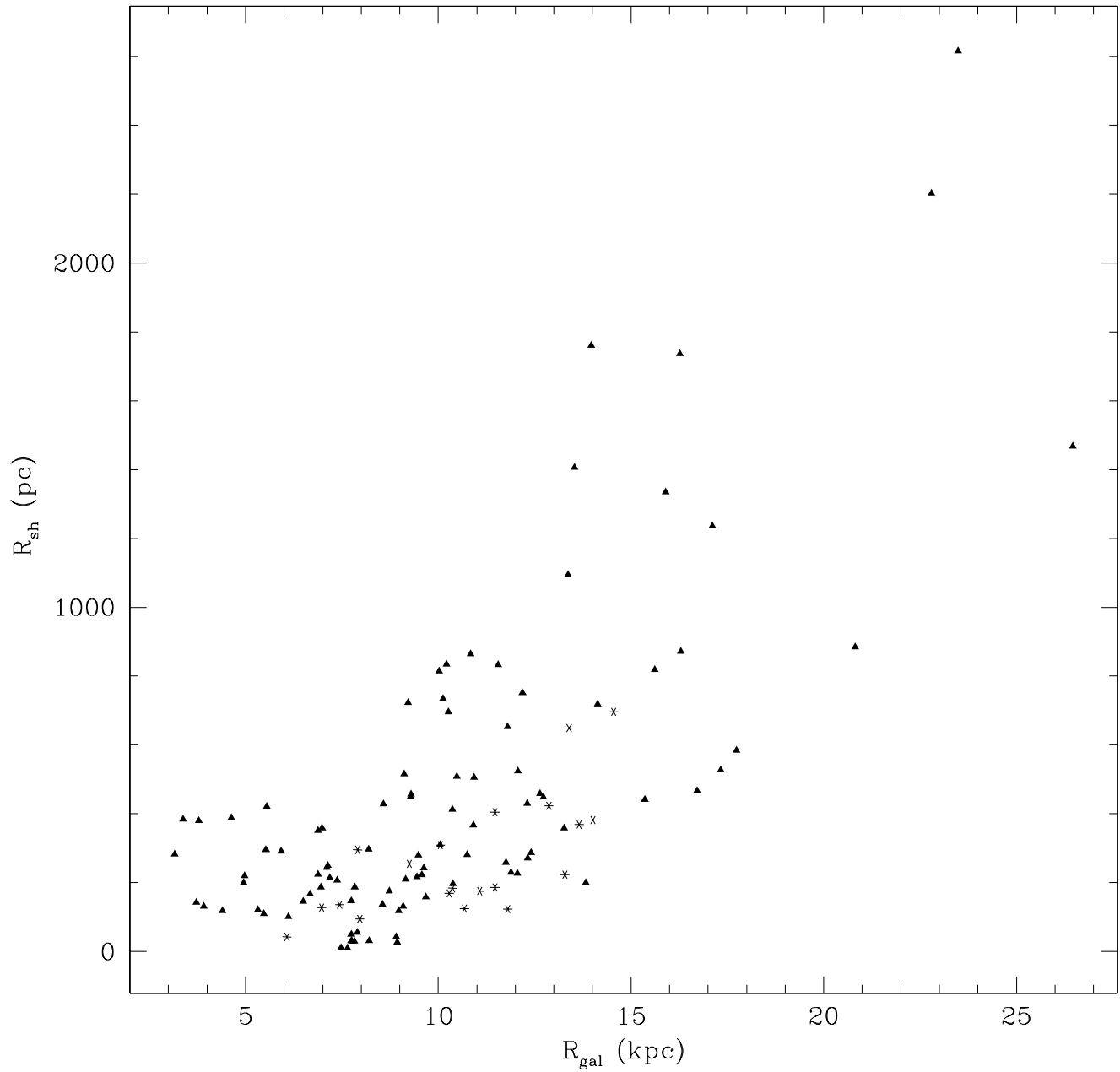}
\caption[Galactic \HI\ shell radius versus galactocentric radius]{Galactic
\HI\ shell radius versus galactocentric radius.  All shells from the SGPS are
plotted with asterisks and shells from \citet{heiles79,heiles84} are plotted
with triangles.  The general trend that large shells are found at large
galactocentric radii is clear.
\label{fig:rgalplot}}
\end{figure}

One objective of the SGPS shell catalog is to explore the Galactic
distribution of shells and determine how their properties vary with Galactic
position.  To this end, we have examined an azimuthally averaged sample of
\HI\ shell size versus Galactic radius.  It has long been noticed in
external galaxies, as well as in the Milky Way, that the largest \HI\ shells
are also at large galactocentric radii
\citep*[e.g.][]{heiles79,deul90,walter99,crosthwaite00}.  \citet{walter99}
noted that in other galaxies, where there are no distance ambiguities that
plague distribution studies of the Milky Way, most large shells are at
galactocentric radii of more than 50\% of the exponential scale length of
the disk.  Using shells from the SGPS and \citet{heiles79,heiles84} we plot
the shell radius as a function of galactocentric radius in
Fig.~\ref{fig:rgalplot}.  Clearly, the largest shells are also at large
Galactic radii.  As mentioned in \S \ref{subsec:biases}, Galactic \HI\ shell
catalogs exhibit a bias towards outer Galaxy shells.  However, the
identification of the same trend in external galaxies is convincing evidence
that the observed effect is real.

A commonly suggested cause of this effect is the increase in the scale
height of the \HI\ disk with galactocentric radius\citep{bruhweiler80}.
Shell size is intimately connected with the \HI\ scale height; as long as a
shell is smaller than the scale height it can expand three-dimensionally.
However, once a shell's size exceeds the scale height expansion in the plane
is largely halted as the shell expands rapidly towards the halo and
eventually bursts, venting its hot interior gas to the halo
\citep{maclow89}.  The correlation between maximum shell size and \HI\ scale
height is best exemplified in dwarf galaxies, such as IC 2574 and Holmberg
II, which have \HI\ disk scale heights of 350-400 and 625 pc respectively
and where \HI\ shells are observed with diameters in excess of 1 kpc
\citep{walter99,puche92}.  In gas-rich spirals, like M31, M33, and the Milky
Way, flaring of the \HI\ disk may allow shells at outer radii to reach very
large sizes while shells at inner radii are confined.  Other causes have
been suggested to explain why the largest shells are at large galactic
radii.  For instance, two-dimensional hydrodynamical simulations by
\citet{wada99} show that \HI\ cavities in the outer disk grow larger than
those in the inner disk because of the disruptive effects of supernovae on
the creation of large-scale structures in the inner disk.  Below we suggest
another possible effect.

The number of large \HI\ shells beyond the solar circle seems to present a
challenge to the theory of shell formation from stellar winds and
supernovae.  Assuming that the energy output of a single O or B type star is
$\sim 10^{51}$ ergs, associations of several hundred massive stars are
needed to create expanding shells with radii of a few hundred parsecs.  It
is difficult to understand how the star formation rates in the outer Galaxy
can be high enough to account for the many large shells.  For example, in
the small SGPS sample there are five shells in the region $267\arcdeg \leq l
\leq 297\arcdeg$, $10.0 \leq R_{gal} \leq 13.3$ kpc with expansion energies
in the range $1.8 - 2.4 \times 10^{53}$ ergs.  That implies a shell surface
density of $\sim 0.3~{\rm kpc^{-2}}$, with each shell requiring more than 150
supernova progenitors to form.  We can compare the observed surface density
with the \citet{mckee97} predictions for the surface density of clusters
with $\mathcal{N}_{\ast h}$ supernova progenitors, as a function of Galactic
radius.  Using their equation (44), scaled to a mean galactocentric radius
of 11 kpc, we find that for $\mathcal{N}_{\ast h} = 150$ the predicted
surface density of clusters is also $\sim 0.3~{\rm kpc^{-2}}$.  Despite this
very good agreement we note that the SGPS catalog is not complete.  The
\citet{heiles84} catalog has a much higher surface density of energetic
shells, which cannot be accounted for by the \citet{mckee97} star formation
rate.  In addition, though the functional form of cluster surface density
allows for populous OB associations at large galactocentric radii, there is
no evidence for any current giant \HII\ regions beyond 11 kpc
\citep*{mckee97,smith78}.  Barring a dramatic change in the star formation
history of the outer Galaxy sometime in the last $\sim 20$ Myr, there seems
to be a discrepancy between the observed giant \HII\ regions and \HI\
supershells.  We are left, therefore, with a population of energetic shells
in the outer Galaxy that cannot be adequately accounted for by the observed
star formation rate.

Attempting to resolve this discrepancy, we consider the possibility that the
expansion energies have been over-estimated.  In particular we explore the
relationship of shell size and expansion energy to the ambient density into
which the shells expand.  If the ambient densities are lower in the outer
Galaxy, as was found around several shells in our catalog, then shells there
should reach larger sizes than those in the inner Galaxy.  Referring to the
expansion energy equation given above, for a fixed input energy and
expansion velocity, the shell radius varies with ambient density, $n_0$, as
$R_{sh} \propto n_0^{-0.36}$.  Unfortunately, this is not a strong
dependency and \HI\ densities do not vary by more than a factor of 3 to 4
across the disk \citep{burton88}.  It is unlikely, then, that shells in the
outer Galaxy grow larger than those in the inner Galaxy by more than a
factor of $\sim 1.7$ (for a density decrease of a factor of 4) due to
external density variation alone.  This does not seem to be a large enough
effect to explain the distribution shown in Fig.~\ref{fig:rgalplot}.
Similarly, the calculated expansion energy is only weekly dependent on the
ambient density, $n_0$, such that $E_E\propto n_0^{1.12}$ so that ambient
density alone cannot account for the highly energetic shells at large
Galactocentric radii.  However, \citet{walter99} point out that the constant
in the expansion energy equation ($5.3 \times 10^{43}$ in the energy
equation given in \S \ref{subsec:props}) is dependent on cooling rates and
includes an assumption about elemental abundances.  Elemental abundances are
not constant throughout the Galaxy and are thought to decrease by more than
an order of magnitude with Galactic radius \citep{shaver83}.  Because metals
are efficient at cooling, a high metallicity environment requires more
energy for a shell to attain a given radius.  Therefore, for shells in the
outer Galaxy where the metallicities are lower, the expansion energies may
be slightly over-estimated.

One can also use a simple timescale argument to explain the observation that
the largest \HI\ shells are found at large Galactic radii.  The frequency
with which a spiral density wave encounters matter in a differentially
rotating disk, expressed in units of $\kappa$, the epicyclic frequency, is
$\nu^2 = m^2(\Omega_p - \Omega)^2 /\kappa^2$, where $m$ is the number of
spiral arms, $\Omega_p$ is the angular velocity of the spiral pattern, and
$\Omega$ is the angular velocity of the gas in the disk.  From the shape of
the spiral arms, the spiral pattern speed is estimated to be in the range
$10$~\kms\ $\leq \Omega_p \leq 20$~${\rm km~s^{-1}~kpc^{-1}}$
\citep[e.g.][]{lepine01,amaral97}.  Therefore at a Galactocentric radius of
$\sim 9$~kpc, the gas and pattern speeds are matched.  Interior to this
corotation radius, $R_{CR}$, the arms move more slowly than the gas,
resulting in a spiral shock on the counter-clockwise side of the arms (as
seen from the North Galactic Pole), or the side towards the Galactic center.
Star formation occurs near this shock and we would expect small \HI\ shells
to form here.  Exterior to the corotation radius the arms move faster than
the disk gas, hence the spiral shock is on the exterior edge of the arms.
At a galactic radius of 8 kpc, the spiral arms travel with a linear
velocity, $v=\Omega_p R_{gal}\approx 120$~\kms (assuming $\Omega_p \sim
15~{\rm km~s^{-1}~kpc^{-1}}$).  At the same distance, however, the gas in
the disk rotates with a velocity of $220$~\kms.  For an arm width of 3~kpc,
the gas crosses the arm in $\sim 30$~Myr, which is comparable to the
formation timescale for a large \HI\ shell \citep{oey97}.  Therefore, if the
first wave of star formation occurs on the leading edge of the spiral arm
then we might expect to see small \HI\ shells formed from several stars
located there.  As the density wave passes through the disk, triggering more
star formation, an \HI\ shell can carve out a progressively larger and
larger volume of the ISM.  This is accompanied by the movement of the spiral
arm such that by the time a large shell has been created, the gas will have
moved away from the spiral arm.

By contrast, at a Galactic radius of 3 kpc, the arms only move at $\sim
45$~\kms, whereas the gas still travels at $220$~\kms.  In this case, the
gas crosses a 3 kpc spiral arm in less than $\sim 17$~Myr, but more
importantly, it also crosses the interarm region in a comparable amount of
time.  Therefore, a shell at a Galactic radius of $< 5$ kpc, which is formed
in a spiral arm will move out of the arm, through the interarm region, and
encounter the next arm before it reaches its maximum radius.  The encounter
with the next spiral shock should be enough to completely disrupt the shell.
This, along with the flaring of the outer \HI\ disk, may explain why large
shells are only seen in the outer Galaxy.  At small Galactic radii they do
not survive long enough to reach a large size.

\subsection{Shells and Spiral Structure}
\label{subsec:shells&spiral}
We can also use the shell catalog to examine how \HI\ shells are spatially
related to Galactic spiral structure.  \citet{thilkerphd} found a strong
correlation between spiral arms and \HI\ shells in NGC 2403, M81 and M101.
In these galaxies many shells are located between the spiral arms, usually
extending from the arms into the interarm regions.  \citet{deul90} noted the
presence of large \HI\ holes in interarm regions of M33.  NGC 4214 has three
large shells between the spiral arms and no detectable shells in the arms
\citep{walter01}.  Similarly, some of the SGPS shells extend into interarm
regions; GSH 277+00+36 and GSH 280+00+59 are particularly clear examples
\citep{mcgriff00}.  A comparison of our shells with the Milky Way spiral
arms is important.  Unfortunately, spiral structure in the Milky Way is
poorly understood.  Our unique position within the Galaxy makes tracing the
spiral structure particularly difficult.  The number of spiral arms,
placement of the arms and even the \HI\ response to the spiral arms are all
largely unknown.  One major obstacle is the necessity of a well understood
rotation curve to calculate kinematic distances.  The Milky Way rotation
curve is uncertain, as are the precise effects of streaming motions.
Molecular line emission and \HII\ regions are often used as tracers of
spiral structure, but many interpretations of these also rely on kinematic
distances.

\begin{figure}
\begin{center}
\plotone{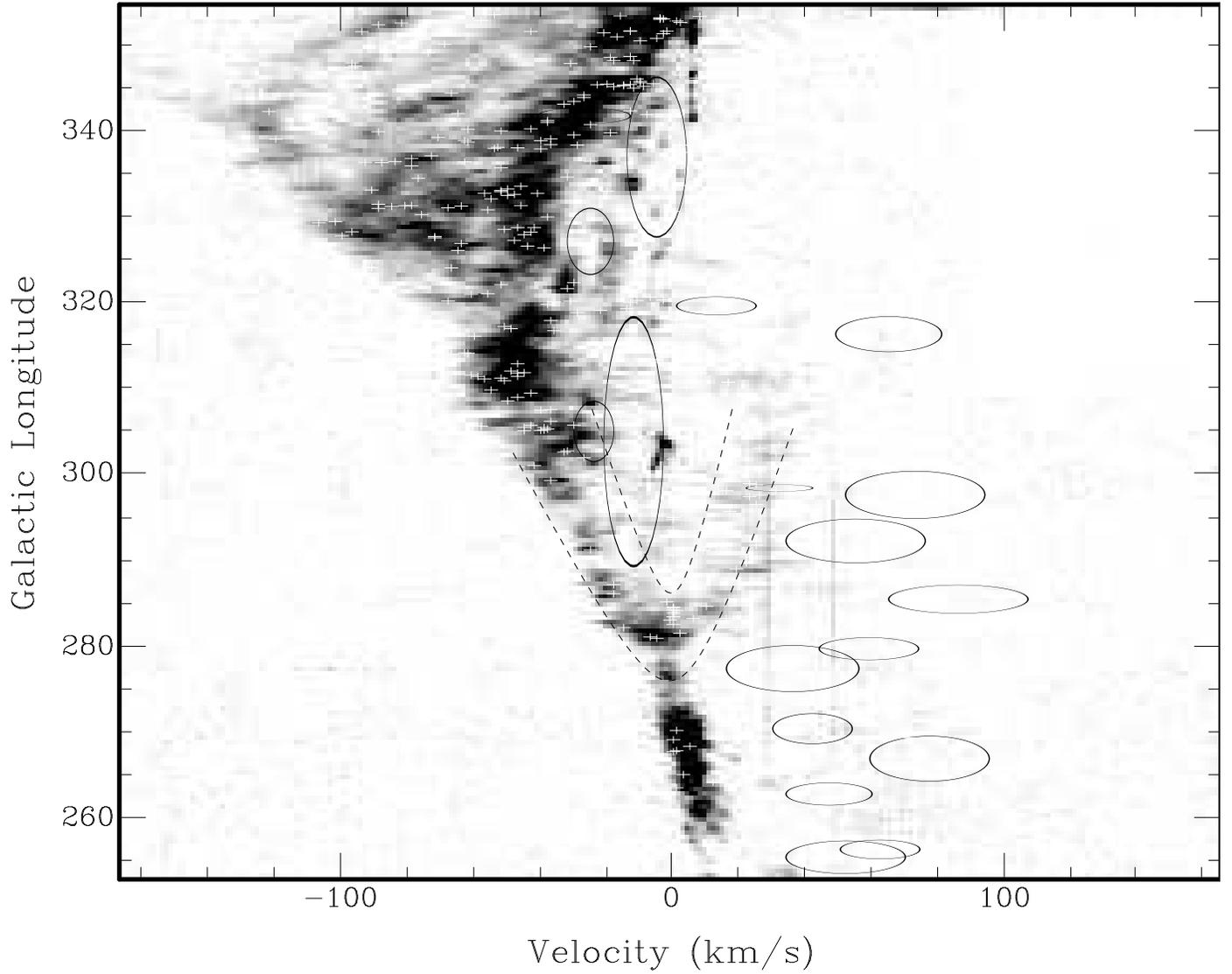}
\caption[SGPS \HI\ shells in {\em l-v} space]{SGPS \HI\ shells plotted on
the $^{12}$CO {\em l-v} image from \cite{dame87}.  The ellipses mark the
SGPS shells and the white crosses mark \HII\ regions cataloged by
\citet{caswell87} on the basis of their recombination lines.  The dashed
lines outline the ``Carina Loop,'' part of the Sagittarius-Carina arm
\label{fig:shell_lv}}
\end{center}
\end{figure}

In order to eliminate the rotation curve dependence it is often preferable
to address Galactic distributions with respect to the longitude-velocity
({\em l-v}) distribution, which does not include any assumptions about
distance.  In Fig.~\ref{fig:shell_lv} we have plotted the SGPS shells on the
$^{12}$CO {\em l-v} diagram from \citet{dame87}.  The CO emission has been
averaged for latitudes over the range $|b|<2\arcdeg$.  Also included on this
figure are \HII\ regions from \citet{caswell87} marked with white crosses.
The shells, marked by ellipses, are located at their cataloged positions and
their plotted sizes are determined by their angular diameters and velocity
widths.  Over the range $v=-100$~\kms\ to $v=0$~\kms\ and $l> 310\arcdeg$
the velocity structure is complicated and consequently the individual spiral
arms are unclear.  The only spiral feature that is immediately
distinguishable is the Carina Loop, part of the Sagittarius-Carina arm near
$l=280\arcdeg$, marked with dashed lines on Fig.\ \ref{fig:shell_lv}.  Few
shells are coincident with the spiral tracers, instead most lie in regions
not associated with spiral structure.  The small chimney-like structure, GSH
298-01+32, is embedded in the edge of this spiral arm, apparently
containing, or perhaps even the result of, an \HII\ region.  Outside of the
loop, towards more positive velocities the shells GSH 277+00+36, GSH
280+00+59, GSH 292-01+55, and GSH 297-00+72 seem to trace the arc of the
loop.  Another interesting feature is the placement of GSH 304-00-12 (large
ellipse at the center of Fig.~\ref{fig:shell_lv}) in an interarm region and
bounded by CO emission from the Coalsack nebula near $v=0$~\kms.

\begin{figure}
\begin{center}
\plotone{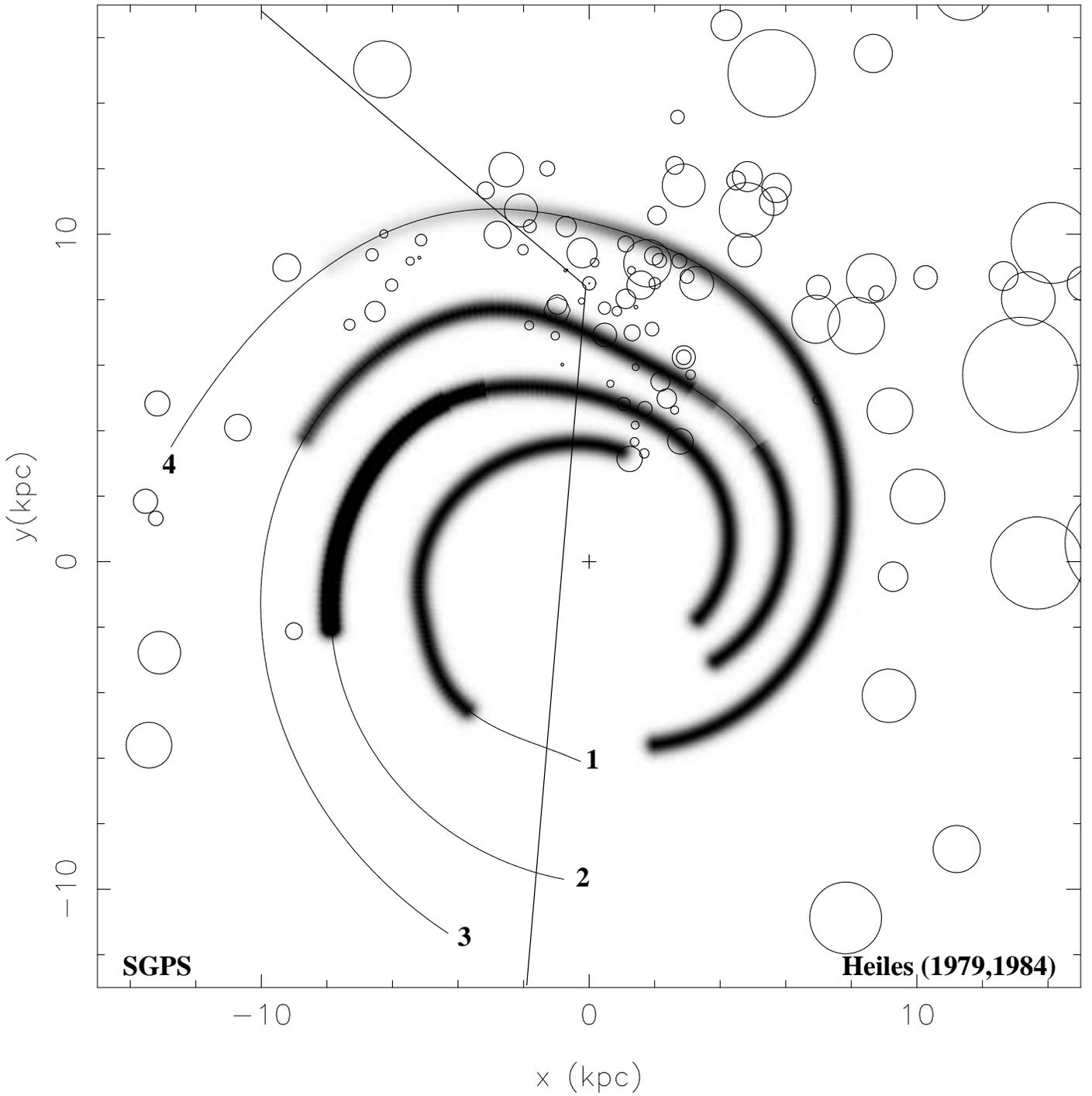}
  \caption[Galactic spiral pattern overlaid with \HI\ shells]{Cataloged
  shells plotted on the spiral pattern of the Galaxy from \citet{taylor93}.
  As marked, shells in the first and second quadrant are from
  \citet{heiles84} and shells in the fourth quadrant are from this work.
  The spiral arms are labelled 1-4 according to the following {\bf 1}:
  Norma, {\bf 2}: Scutum-Crux, {\bf 3}: Sagittarius-Carina, and {\bf 4}:
  Perseus.  The Galactic center is marked by a cross-hair and the position of
  the Sun is marked by circle with a dot at the center.  The SGPS survey
  region is marked by straight, solid lines.
\label{fig:shelldist}}
\end{center}
\end{figure}

To expand the distribution study to more global scales, we have included the
shells with $|b| < 5\arcdeg$ from \citet{heiles79,heiles84}.
Figure~\ref{fig:shelldist} is a diagram of \HI\ shells from the SGPS and
\citet{heiles79,heiles84} catalogs plotted on the spiral arms of the Galaxy
as defined in the \citet{taylor93} electron density model.  The
\citet{taylor93} model fits the arms on the basis of the \citet{georgelin76}
spiral model from \HII\ regions and radio-continuum features that mark the
spiral arm tangents.  The model describes the electron density distribution
of the Galaxy, which is very much related to star formation, and is the most
detailed model of the spiral arm positions available.  The arms given by
\citet{taylor93} have been extended (narrow lines on
Fig.~\ref{fig:shelldist}) to approximate a spiral pattern.  It should be
stressed that the extensions are extrapolations and are not based on spiral
tracers.  Shell positions are also uncertain.  Departures from circular
rotation caused either by the underlying spiral pattern or other systemic
effects are not accounted for in their positions. These departures can be as
high as $\sim 10$~\kms, and therefore the positions of all shells are
uncertain at the 5-20\% level.  The \citet{heiles79,heiles84} shells are on
the right-hand side of the figure and the shells on the left are from the
SGPS.  Many of the large cluster of local shells from the
\citet{heiles79,heiles84} catalog seem to lie between the Sagittarius-Carina
arm and the Perseus arm.  Among the SGPS shells, some are clearly located
between the spiral arms.  The extreme distance of several of the shells,
however, makes it difficult to compare them to the spiral arms because they
are beyond the known extent of the arms.

Figs.~\ref{fig:shell_lv} \& \ref{fig:shelldist} seem to suggest that an
appreciable number of \HI\ shells overlap interarm regions.  The most
convincing examples of interarm shells are the outer Galaxy chimneys, GSH
277+0+36 and GSH 280+0+59, and the Coalsack shells, GSH 305+01-24 and GSH
304+00-12, which are all found on the edge of the Sagittarius-Carina spiral
arm and open into the region between the Sagittarius-Carina and Perseus
arms.  Because only a portion of one spiral arm is clear in the {\em l-v}
diagram and the distribution of shells from a face-on view of the Galaxy is
subject to distance determination uncertainties, it is not possible to
determine how strongly the correlation holds.  We note that the distribution
could be partially a selection effect; \HI\ shells at the edge of spiral
arms are easier to detect than those in arms.  Certainly there are enough
shells between spiral arms to raise the question, why are they there?

While we expect star formation to occur in the spiral arms, the $\sim 30$
Myr required to form a large shell is long enough that shells may migrate
out of the spiral arms into interarm regions.  So, this distribution is not
surprising.  Another important component to the shell distribution may be
the density structure of the Galactic disk, in particular the density
gradient at the transition from spiral arm to interarm region.  At this
transition the density gradient may lead to enhanced expansion of the shell.
Unlike the leading edge of the arm, the gravitational potential well at the
back side of the arm is not very steep and there are no shocks to disturb
the shell's expansion.  As a result, the shell may largely escape the spiral
arm and expand along the density gradient into the interarm region.  The
result could be inordinately large shells in interarm regions.

\subsubsection{Simple Spiral Structure Model}
\label{subsubsec:spiralmod}
To explore this theory we need to know the strength of the density gradient,
which can be naively estimated from a simple spiral perturbation model of
the disk.  Following the arguments of \citet[][hereafter
RH84]{roberts84}, they define the two-dimensional, spirally perturbed disk
gravitational potential as
\begin{equation}
U(r,\theta,t) = U_0(r) \left\{ 1 + \xi(r) \cos \left[ m \theta - m \Omega_p
t +\Phi(r) \right] \right\},
\label{eq:modelspir}
\end{equation}
where $r$ is the galactocentric radius, $\theta$ is the azimuthal angle,
$U_0$ is the unperturbed potential, $\Omega_p$ is the pattern speed, and
$\xi$ and $\Phi$ are the amplitude and phase, respectively, of the spiral
perturbation. The unperturbed potential, $U_0$, is that of a Toomre disk
\citep{toomre63} given by
\begin{equation}
U_0\left(r\right) = - \frac{B^2 a^3}{\left(a^2 + r^2\right)^{1/2}},
\label{eq:toomre}
\end{equation}
where the constants $a = 7$~kpc and $B=0.0576~{\rm Myr^{-1}}$ have been
chosen to give a peak circular velocity of 220~\kms\ at 8.5 kpc.  The
amplitude of the spiral perturbation is,
\begin{equation}
\xi(r) = \frac{A}{5} \, \frac{a^2 r^2}{\left(a^2 + r^2\right)^2}.
\label{eq:spiralamp}
\end{equation}     
The constant, $A=0.067$, was chosen to make a spiral perturbing force that
is $5-10$\% of the unperturbed force.  The phase of perturbation is given by
\begin{equation}
\Phi (r) = \frac{2 \ln \left[1+\left(r/r_0\right)^j \right]}{j\tan i_0},
\label{eq:spiralphase}
\end{equation}
where $i_0$ is the pitch angle of the spiral, set to 10\arcdeg\ in RH84.
The phase is defined to make a transition between a bar-like potential at
$r<r_0 = 1$ kpc and a spiral-like potential at $r>r_0$.  The power $j=5$
determines how sharply this transition takes place.

We are concerned with the surface density response to the gravitational
potential. Therefore Poisson's equation must be solved for the surface
density from the Toomre and spiral potentials.  Given the unperturbed
potential in equation~\ref{eq:toomre}, the unperturbed surface density from
RH84 is:
\begin{equation}
\sigma_o \left(r\right) = \frac{B^2 a^4}{2\pi G\left(a^2 + r^2\right)^{3/2}}.
\end{equation}
Following \citet{lin69} and \citet{freeman75} we solve Poisson's equation
with the spiral potential for the linear density response of the total gas
and stars, by making an asymptotic approximation that the pitch angle is
small and therefore the quantity $[\Phi^{\prime}(r)r]^{-1}$ is small.  The
total surface density perturbation (gas and stars) is then given by:
\begin{equation}
\sigma_p = \frac{U_o(r) \xi (r)}{2 \pi G} \Phi^{\prime}(r) \cos [m\theta
-m\Omega_p t + \Phi(r)],
\label{eq:spirdens}
\end{equation}
where $m$ is the number of arms.  The total surface density, including the
Toomre disk, is simply $\sigma_t = \sigma_o(r) + \sigma_p(r)$.  Using the
two-dimensional surface density given by these equations we modified the
original RH84 parameters to better fit the \citet{georgelin76} and
\citet{taylor93} model for the spiral arms.  Rather than assuming a
two-armed spiral we used a four-arm spiral with a pitch angle of $11\fdg5$
and decreased the amplitude of the spiral perturbations so that $A=0.05$.
We found that these modifications gave a better fit to the \citet{taylor93}
model for the spiral pattern and agree well with recent work by
\citet{vallee02}. We note, though, that the model employed here is not
designed to perfectly fit the observed Milky Way, but rather given what we
do know about the spiral structure near to the Sun, to estimate the relative
magnitudes of the $z$ and $r$ components of the density gradient.

The question of how the \HI\ reacts to the spiral potential is non-trivial.
The density derived in equation~\ref{eq:spirdens} neglects the effects of
star formation, which through photo-dissociation may be significant in
minimizing the arm-interarm density contrasts in \HI.  In addition, while
the stellar disk dominates the mass density at small galactic radii, the gas
disk dominates over the stellar disk for large radii.  We have therefore
modified the surface density to reflect the smaller arm-interarm contrasts
seen in the Galaxy \citep{burton88} and to decrease the axisymmetric drop in
density.  We relate the surface density to the volume density at midplane
according to $\Sigma_{HI} = \sqrt{2\pi}\sigma_z\rho_{HI}(0)$, where
$\sigma_z = 120$ pc \citep{dickey90}. Finally, the parameters have been
adjusted to give a fiducial density of $n_{HI} \approx 0.3~{\rm cm^{-3}}$ at
the position of the Sun. 

\subsubsection{Comparison of the Density Gradient}
\label{subsubsec:gradvec}
As suggested above, the ambient medium has an effect on the radius and
energetics of a supershell. It has been shown that a shell expanding in a
stratified medium, i.e.\ from the Galactic disk into the halo, will
experience exaggerated expansion along the density gradient, forming a
chimney \citep[e.g.][]{maclow88}.  Though magnetic fields in the plane can
confine a shell for $\sim 20$~Myr \citep{tomisaka98}, models show that the
effect of the density gradient will be enough for a large shell to
eventually blow out of the plane.  We hypothesize that if the density
gradient from arm to interarm regions is strong enough the same effect can
form an in-plane chimney.  We therefore need to understand the density
gradient around spiral arms in order to assess the effect of the spiral arms
and interarm regions on shells.

\begin{figure}
\centering
\epsscale{0.65}
\plotone{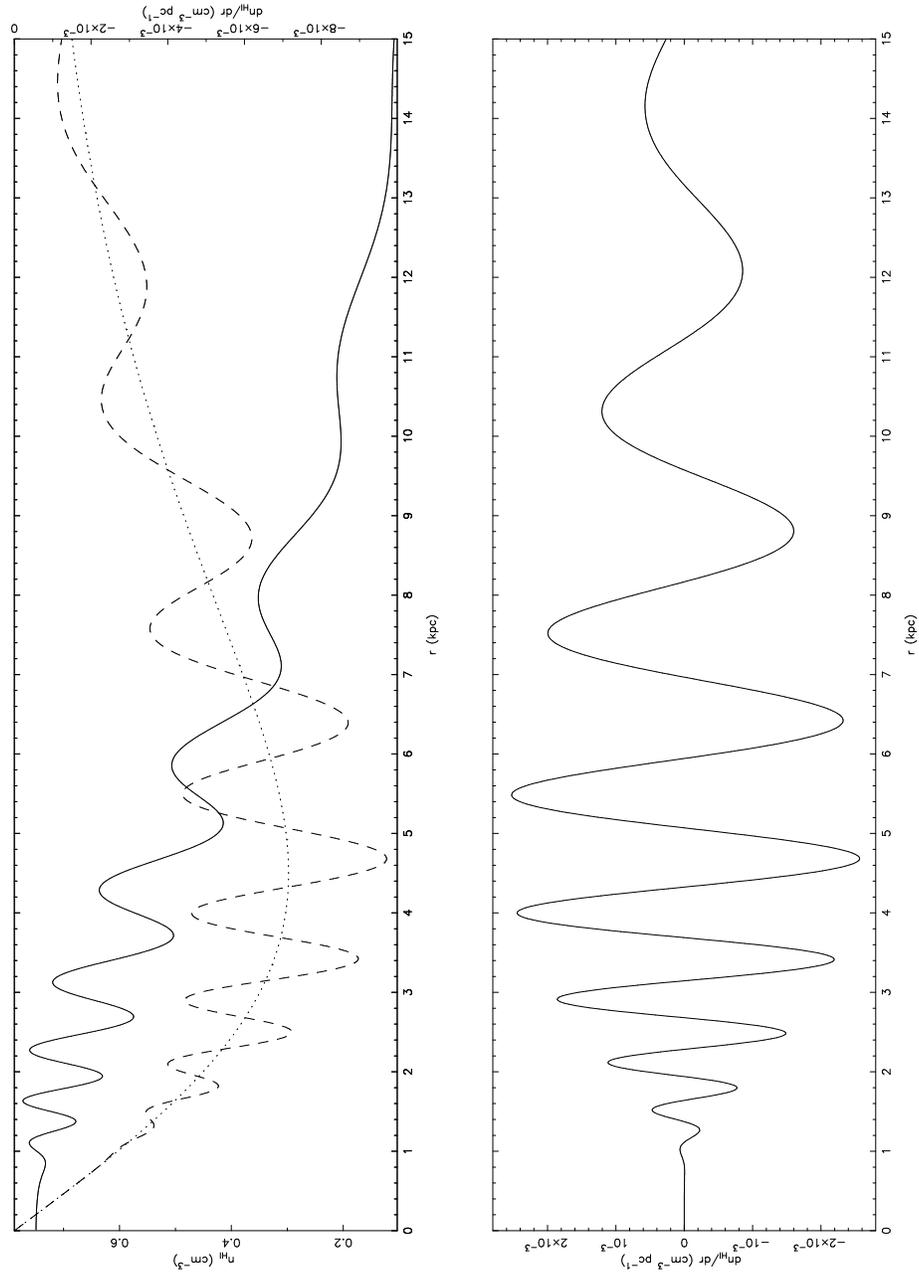}
\caption[Profiles of \HI\ number density and density gradient as a function
of $R_{\rm gal}$] {Modeled \HI\ density and density gradient along the line of sight
crossing the Sun and the Galactic center.  The top panel shows the
\HI\ density profile, $n_{HI}(r)$, with a solid line.  The dashed line is the
one-dimensional density gradient profile, $dn_{HI}(r)/dr$, and the dotted line is the
density gradient for an axisymmetric potential, not including the spiral
perturbation.  The scale for the volume density is at the left and the scale
for the density gradient is shown on the right.  The bottom panel is the
residual density gradient obtained by subtracting the axisymmetric component
from the total density gradient.  This plot shows the effect of the spiral
arms alone.
\label{fig:denspro}}
\end{figure}

\begin{figure}
\centering
\includegraphics[angle=-90,width={\textwidth}]{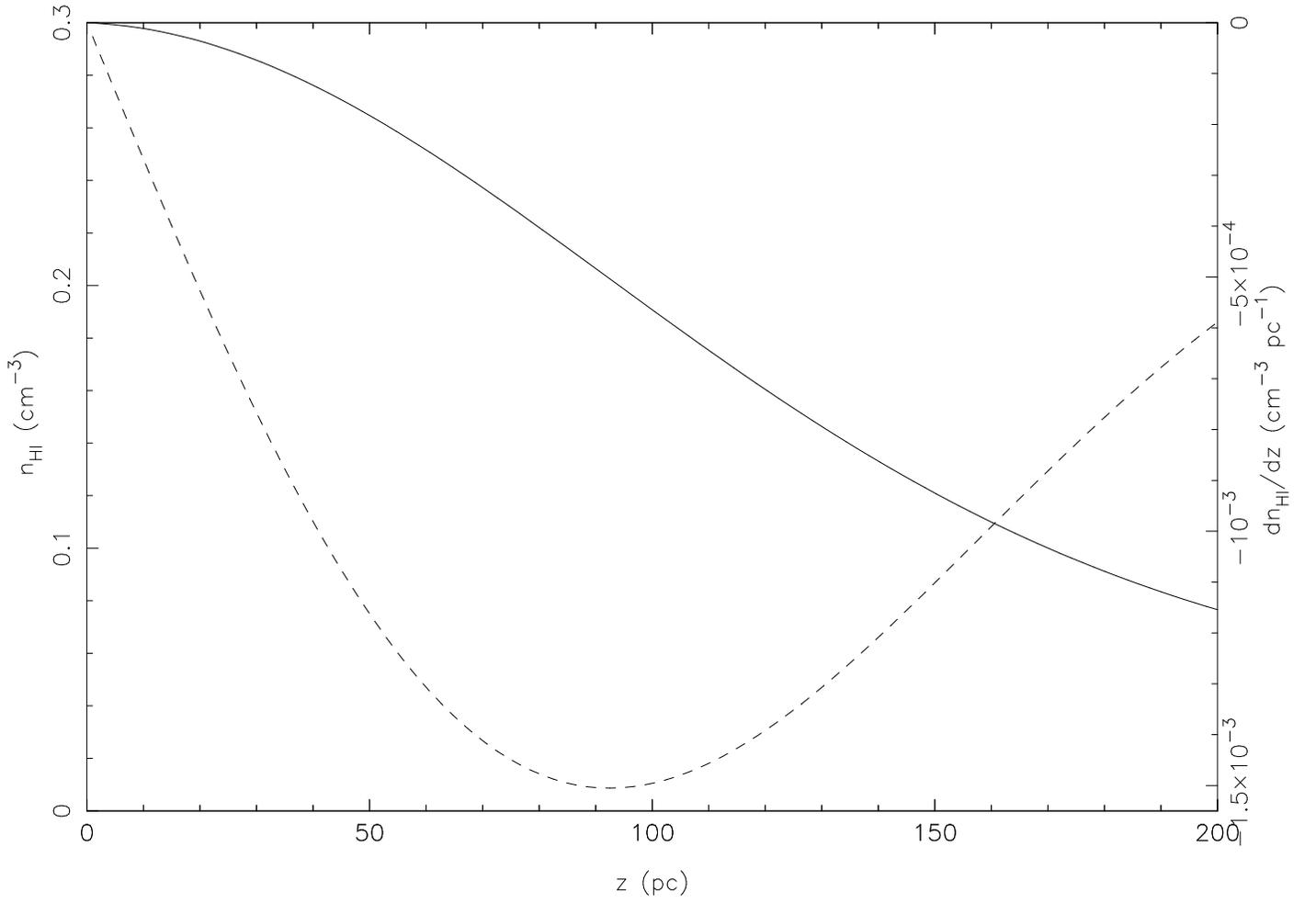}
\caption[Profile of \HI\ number density and density gradient as a function of $z$]
{\HI\ density and density gradient as a function of height above the
galactic plane, $z$.  The solid line is the \HI\ number density with the
axis marked to the left.  The dashed line is the density gradient,
$dn_{HI}/dz$, with the axis marked on the right.
\label{fig:zdens}}
\end{figure}

Figures~\ref{fig:denspro} and \ref{fig:zdens} show radial and $z$ profiles
of \HI\ number density and density gradient.  The radial profile was taken
along the line-of-sight from the Sun to the Galactic center and the $z$
profile starts at mid-plane and extends to $z=200$ pc.  Obviously, both
models grossly oversimplify the Galactic disk, but they are instructive to
show how the density varies across the disk and out of the plane.  Also
shown in Fig.~\ref{fig:denspro} is a density gradient calculated from just
the spiral perturbation to the density, but not including the underlying
axisymmetric disk.  This should allow for a more direct comparison of the
gradient around spiral arms with that out of the plane.  Comparing the two
gradients, we find that for this model, at a typical height of $z =100$ pc
above the mid-plane, the density gradients are of comparable magnitude.
Assuming that the interstellar pressure is largely determined by the
density, an expanding shell close to the Galactic plane should feel a
stronger gradient away from the arm than out of the plane.  Consequently the
shell should expand rapidly away from the arm through the interarm medium.
Towards the arm, expansion will be impeded by the higher density.  As the
shell continues to expand vertically it will reach a $z$-height of $\sim
100$ pc, at which point the density gradient out of the plane will begin to
dominate.  

\subsection{Discussion}
\label{subsec:shelldisc}
This simple model presents a possible explanation for the observation that
many \HI\ shells in the Galaxy are between the spiral arms.  The combined
effects of the density gradient and migration of shells may lead to a
spatial offset of shells away from the spiral shock and into the interarm
regions.  Because shells expanding into interarm regions can reach
exaggerated sizes, their formation energy requirements are reduced to levels
more consistent with star formation rates.  Other environmental factors that
influence shell evolution may enhance the effect suggested here, namely the
magnetic field and cosmic ray pressures.  These may also show strong
arm-interarm gradients, but it is beyond the scope of this paper to model
them.

We note that the density gradient determined here depends on the number
of arms and the pitch angle of the spiral.  The model employed here has four
arms with a relatively small pitch angle.  These quantities, however, are
still not well determined.  We find that for a much less tightly wound
spiral, with a pitch angle of $20\arcdeg$, the maximum density gradient is a
factor of two less than that for a pitch angle of $11\fdg5$.  In this case,
the density gradient away from spiral arms would never dominate, but is
still comparable to the density gradient out of the disk.  So, while the
model developed in \S \ref{subsubsec:spiralmod} was based on previous
estimates of Milky Way parameters (e.g.\ RH84, Vall\'{e}e 2002), it is
sufficiently robust that its specifics can be modified and the main result,
that the $z$ and $r$ components of the density gradient are of comparable
magnitude, will hold.  This fact should be considered when examining
external galaxies in which the spiral structure may be significantly
different than in the Milky Way.

A similar observation of shell-like structures at the edge of spiral arms
was made by \citet{crosthwaite00}, who came to a different conclusion.
\citet{crosthwaite00} examined the \HI\ in the spiral galaxy IC 342 and
found several large, shell-like structures in the outer galaxy ($R>
R_{CR}$).  In particular, the question of whether the fine structure in the
spiral arms is due to shells or flocculent arms is addressed.  They
determine that the holes do not have the kinematic signatures of shells,
that they are not necessarily associated with star formation in the form of
H$\alpha$ emission or \HII\ regions, that the inferred expansion energies
are too large ($E_{E}\sim 3\times 10^{53}$ ergs) to be shells, and most
importantly, they argue that the shells do not appear to be experiencing
shearing due to differential rotation.  From this last point they argue that
shearing would have significantly affected their observed shells in the
timescales required to create holes that large with stellar winds and
supernovae.  They instead explore the possibilities that the structures were
formed from negative shear that created spurs on the gas arms, or
alternately that the flocculent structure is due to gas instabilities that
created shell-like structures and holes, similar to the instabilities
suggested by \citet{wada99}.  They determine that the surface density
enhancements in the spiral arm segments are insufficient to promote negative
shear and hence perpendicular arm-spurs.  They therefore conclude that the
data supports the hypothesis that the shell-like structures are formed from
two-fluid gravitational instabilities in the gas disk, not shells.  We
cannot rule out that shells seen at the edges of arms in the Milky Way are
also due to gravitational instabilities.  However, in our limited sample of
SGPS shells, the kinematic signatures are distinctly characteristic of
expanding structures.  Also the structure of the shell walls is consistent
with shell walls formed by compression.  GSH 277+00+36, for example, exhibits
a very sharp shell wall and is cohesive over more than 40~\kms\ of velocity
dispersion.  Both characteristics seem improbable for an object formed by
gravitational instability.

The suggested scenario may explain the position and formation energy of
supershell GSH 277+00+36 \citep{mcgriff00}.  In the early stages of its
evolution the shell may have expanded away from the spiral arm through a
region where the density steadily decreased from its central position.  In
this way it may have attained an exceptionally large size and therefore may
not have required $10^{53}$ ergs to form.  At about 100 pc, however, the
gradient out of the plane would have begun to dominate and the shell
expanded quickly into the halo forming a chimney.  The rapid expansion and
pressure differential between the halo gas and the pressurized shell
interior may have resulted in Rayleigh-Taylor instabilities that formed the
chimney channels seen extending from the shell to the halo.  If this
scenario is correct, we may see similar, though perhaps less developed,
Rayleigh-Taylor instabilities developing in the walls perpendicular to the
plane.

The idea that the number of large, energetic shells observed in galaxies is
irreconcilable with star formation is not a new one
\cite[e.g.][]{rhode99,crosthwaite00}.  Numerous alternative formation
theories have been developed to overcome this ``energy crisis,'' including the
impact of high velocity clouds \citep{tenorio87}, gamma-ray bursts
\citep{loeb98}, and gravitational or thermal ISM instabilities
\citep{wada00}.  Though the alternatives are appealing, and may hold true
for some shells, it seems clear that we do not yet understand shells well
enough to resolve the crisis.  How are there so many energetic shells?
Though the energy requirements can be lessened somewhat by allowing for
shells expanding into interarm regions, alternative theories will have to be
invoked for some shells.
\section{Conclusions}
\label{sec:catconcl}
We have discovered nineteen new \HI\ shells in the Southern Galactic Plane
Survey.  These shells vary in radius from 40 pc to 700 pc, have expansion
velocities between 6~\kms\ and 20~\kms, have expansion energies between
$10^{51}$ ergs and $10^{53}$ ergs and are distributed throughout the fourth
quadrant of the Galaxy.  We have used this new catalog, along with those of
\citet{heiles79,heiles84}, to examine the distribution of large \HI\ shells
in the Milky Way.  We find a tendency for large, energetic shells to be
located at large Galactocentric radii; a tendency that has also been
observed in external galaxies \citep[e.g.][]{deul90,walter99}.  We also show
that many shells are located between the spiral arms of the Galaxy.  We have
used basic timescale arguments to show that the timescales are such that gas
moves out of a spiral arm in a time that is comparable to the time required
for the formation of a large ($R>100$ pc) shell.  Hence, we believe that
large shells will often end their lives between spiral arms.  We have used
the same timescale arguments to show that in the inner Galaxy, where few
large shells are observed, a shell would move out of an arm, through the
interarm region, and be struck by the next spiral arm before it could grow
very large.  This may explain why large shells are only seen at large
Galactic radii.

We have also used spiral density wave theory to explore the radial density
structure of the Galactic disk as a result of spiral arms.  We compared this
to the density structure extending out of the Galactic plane towards the
halo to show that the density gradient away from spiral arms is comparable
to that from the disk to the halo.  Because simulations have shown that
shells expanding out of the plane tend to elongate in that direction, we
suggest an analogous scenario in which a shell experiences runaway expansion
away from spiral arms.  This effect, combined with the multiple generations of
star formation required to create large shells, should lead to a population
of interarm shells.

\acknowledgements This research was supported by NSF grant AST-9732695 to
the University of Minnesota and NASA Graduate Student Researchers Program
(GSRP) Fellowship NGT 5-50250 to N.\ M.\ M.-G..  B.\ M.\ G.\ acknowledges
the support of a Clay Fellowship awarded by the Harvard Smithsonian Center
for Astrophysics.  We would like to thank D.\ Johnstone and R.\ Benjamin for
fruitful discussions, K.\ Barnard for help compiling the catalog, and R.\
Wark and J.\ Reynolds for observational support throughout the SGPS.



\small 
\begin{thebibliography}{57}
\expandafter\ifx\csname natexlab\endcsname\relax\def\natexlab#1{#1}\fi

\bibitem[{Amaral \& L\`epine(1997)}]{amaral97}
Amaral, L.~H. \& L\`epine, J. R.~D. 1997, \mnras, 286, 885

\bibitem[{{Barnes, D.\ G. et al.}(2001)}]{barnes01}
{Barnes, D.\ G. et al.} 2001, \mnras, 322, 486

\bibitem[{Bruhweiler {et~al.}(1980)Bruhweiler, Gull, \& Sofia}]{bruhweiler80}
Bruhweiler, F.~C., Gull, Theodore R. abd~Kafatos, M., \& Sofia, S. 1980, \apjl,
  238, L27

\bibitem[{Burton(1988)}]{burton88}
Burton, W.~B. 1988, Galactic and Extragalactic Radio Astronomy, 2nd edn.
  (Springer), 295

\bibitem[{Caswell \& Haynes(1987)}]{caswell87}
Caswell, J.~L. \& Haynes, R.~F. 1987, \aap, 171, 261

\bibitem[{{Chevalier}(1974)}]{chevalier74}
{Chevalier}, R.~A. 1974, \apj, 188, 501

\bibitem[{{Cioffi} {et~al.}(1988){Cioffi}, {McKee}, \&
  {Bertschinger}}]{cioffi88}
{Cioffi}, D.~F., {McKee}, C.~F., \& {Bertschinger}, E. 1988, \apj, 334, 252

\bibitem[{Crosthwaite {et~al.}(2000)Crosthwaite, Turner, \& Ho}]{crosthwaite00}
Crosthwaite, L.~P., Turner, J.~L., \& Ho, P. T.~P. 2000, \aj, 119, 1720

\bibitem[{{Dame} {et~al.}(1987){Dame}, {Ungerechts}, {Cohen}, {de Geus},
  {Grenier}, {May}, {Murphy}, {Nyman}, \& {Thaddeus}}]{dame87}
{Dame}, T.~M., {Ungerechts}, H., {Cohen}, R.~S., {de Geus}, E.~J., {Grenier},
  I.~A., {May}, J., {Murphy}, D.~C., {Nyman}, L.~A., \& {Thaddeus}, P. 1987,
  \apj, 322, 706

\bibitem[{de~Blok \& Walter(2000)}]{deblok00}
de~Blok, W. J.~G. \& Walter, F. 2000, \apjl, 537, L95

\bibitem[{{Deul} \& {den Hartog}(1990)}]{deul90}
{Deul}, E.~R. \& {den Hartog}, R.~H. 1990, \aap, 229, 362

\bibitem[{{Dickey} \& {Lockman}(1990)}]{dickey90}
{Dickey}, J.~M. \& {Lockman}, F.~J. 1990, \araa, 28, 215

\bibitem[{{Ehlerov\'a} \& {Palou\v{s}}(1996)}]{ehlerova96}
{Ehlerov\'a}, S. \& {Palou\v{s}}, J. 1996, \aap, 313, 478

\bibitem[{{Fich} {et~al.}(1989){Fich}, {Blitz}, \& {Stark}}]{fich89}
{Fich}, M., {Blitz}, L., \& {Stark}, A.~A. 1989, \apj, 342, 272

\bibitem[{Freeman(1975)}]{freeman75}
Freeman, K.~C. 1975, Stars and Stellar Systems, Vol.~VI, Galaxies and the
  Universe (The University of Chicago Press), 409

\bibitem[{{Georgelin} \& {Georgelin}(1976)}]{georgelin76}
{Georgelin}, Y.~M. \& {Georgelin}, Y.~P. 1976, \aap, 49, 57

\bibitem[{{Heiles}(1979)}]{heiles79}
{Heiles}, C. 1979, \apj, 229, 533

\bibitem[{{Heiles}(1984)}]{heiles84}
---. 1984, \apjs, 55, 585

\bibitem[{{Heiles}(1998)}]{heiles98}
---. 1998, \apj, 498, 689

\bibitem[{Kim(1998)}]{kimphd}
Kim, S. 1998, PhD thesis, The Australian National University

\bibitem[{{Kim} {et~al.}(1998){Kim}, {Staveley-Smith}, {Dopita}, {Freeman},
  {Sault}, {Kesteven}, \& {McConnell}}]{kim98}
{Kim}, S., {Staveley-Smith}, L., {Dopita}, M.~A., {Freeman}, K.~C., {Sault},
  R.~J., {Kesteven}, M.~J., \& {McConnell}, D. 1998, \apj, 503, 674

\bibitem[{{L{\' e}pine} {et~al.}(2001){L{\' e}pine}, {Mishurov}, \&
  {Dedikov}}]{lepine01}
{L{\' e}pine}, J.~R.~D., {Mishurov}, Y.~N., \& {Dedikov}, S.~Y. 2001, \apj,
  546, 234

\bibitem[{{Lin} {et~al.}(1969){Lin}, {Yuan}, \& {Shu}}]{lin69}
{Lin}, C.~C., {Yuan}, C., \& {Shu}, F.~H. 1969, \apj, 155, 721

\bibitem[{{Loeb} \& {Perna}(1998)}]{loeb98}
{Loeb}, A. \& {Perna}, R. 1998, \apjl, 503, L35

\bibitem[{{Mac Low} \& {McCray}(1988)}]{maclow88}
{Mac Low}, M. \& {McCray}, R. 1988, \apj, 324, 776

\bibitem[{{Mac Low} {et~al.}(1989){Mac Low}, {McCray}, \& {Norman}}]{maclow89}
{Mac Low}, M., {McCray}, R., \& {Norman}, M.~L. 1989, \apj, 337, 141

\bibitem[{{Maciejewski} {et~al.}(1996){Maciejewski}, {Murphy}, {Lockman}, \&
  {Savage}}]{maciejewski96}
{Maciejewski}, W., {Murphy}, E.~M., {Lockman}, F.~J., \& {Savage}, B.~D. 1996,
  \apj, 469, 238

\bibitem[{{Mashchenko} {et~al.}(1999){Mashchenko}, {Thilker}, \&
  {Braun}}]{mashchenko99}
{Mashchenko}, S.~Y., {Thilker}, D.~A., \& {Braun}, R. 1999, \aap, 343, 352

\bibitem[{{McClure-Griffiths} {et~al.}(2001{\natexlab{a}}){McClure-Griffiths},
  {Dickey}, {Gaensler}, \& {Green}}]{mcgriff01c}
{McClure-Griffiths}, N.~M., {Dickey}, J.~M., {Gaensler}, B.~M., \& {Green},
  A.~J. 2001{\natexlab{a}}, \apj, 562, 424

\bibitem[{{McClure-Griffiths} {et~al.}(2000){McClure-Griffiths}, {Dickey},
  {Gaensler}, {Green}, {Haynes}, \& {Wieringa}}]{mcgriff00}
{McClure-Griffiths}, N.~M., {Dickey}, J.~M., {Gaensler}, B.~M., {Green}, A.~J.,
  {Haynes}, R.~F., \& {Wieringa}, M.~H. 2000, \aj, 119, 2828

\bibitem[{{McClure-Griffiths} {et~al.}(2001{\natexlab{b}}){McClure-Griffiths},
  {Green}, {Dickey}, {Gaensler}, {Haynes}, \& {Wieringa}}]{mcgriff01a}
{McClure-Griffiths}, N.~M., {Green}, A.~J., {Dickey}, J.~M., {Gaensler}, B.~M.,
  {Haynes}, R.~F., \& {Wieringa}, M.~H. 2001{\natexlab{b}}, \apj, 551, 394

\bibitem[{McCray \& Kafatos(1987)}]{mccray87}
McCray, R. \& Kafatos, M. 1987, \apj, 317, 190

\bibitem[{{McKee} \& {Williams}(1997)}]{mckee97}
{McKee}, C.~F. \& {Williams}, J.~P. 1997, \apj, 476, 144

\bibitem[{Mintner {et~al.}(2001)Mintner, Lockman, Langston, \&
  Lockman}]{mintner01}
Mintner, A.~H., Lockman, F.~J., Langston, G.~I., \& Lockman, J.~A. 2001, \apj,
  555, 868

\bibitem[{Oey \& Clarke(1997)}]{oey97}
Oey, M.~S. \& Clarke, C.~J. 1997, \mnras, 289, 570

\bibitem[{{Puche} {et~al.}(1992){Puche}, {Westpfahl}, {Brinks}, \&
  {Roy}}]{puche92}
{Puche}, D., {Westpfahl}, D., {Brinks}, E., \& {Roy}, J.-R. 1992, \aj, 103,
  1841

\bibitem[{{Rand} \& {Stone}(1996)}]{rand96}
{Rand}, R.~J. \& {Stone}, J.~M. 1996, \aj, 111, 190

\bibitem[{Rhode {et~al.}(1999)Rhode, Salzer, Westpfahl, \& Radice}]{rhode99}
Rhode, K.~L., Salzer, J.~J., Westpfahl, D.~J., \& Radice, L.~A. 1999, \aj, 118,
  323

\bibitem[{{Roberts} \& Hausman(1984)}]{roberts84}
{Roberts}, W.~W. \& Hausman, M.~A. 1984, \apj, 277, 744

\bibitem[{{Shaver} {et~al.}(1983){Shaver}, {McGee}, {Newton}, {Danks}, \&
  {Pottasch}}]{shaver83}
{Shaver}, P.~A., {McGee}, R.~X., {Newton}, L.~M., {Danks}, A.~C., \&
  {Pottasch}, S.~R. 1983, \mnras, 204, 53

\bibitem[{{Smith} {et~al.}(1978){Smith}, {Biermann}, \& {Mezger}}]{smith78}
{Smith}, L.~F., {Biermann}, P., \& {Mezger}, P.~G. 1978, \aap, 66, 65

\bibitem[{{Spangler}(2001)}]{spangler01}
{Spangler}, S.~R. 2001, Space Science Reviews, 99, 261

\bibitem[{{Stanimirovi\'{c}} {et~al.}(1999){Stanimirovi\'{c}},
  {Staveley-Smith}, {Dickey}, {Sault}, \& {Snowden}}]{stanimirovic99}
{Stanimirovi\'{c}}, S., {Staveley-Smith}, L., {Dickey}, J.~M., {Sault}, R.~J.,
  \& {Snowden}, S.~L. 1999, \mnras, 302, 417

\bibitem[{{Staveley-Smith} {et~al.}(1997){Staveley-Smith}, {Sault},
  {Hatzidimitriou}, {Kesteven}, \& {McConnell}}]{staveley-smith97}
{Staveley-Smith}, L., {Sault}, R.~J., {Hatzidimitriou}, D., {Kesteven}, M.~J.,
  \& {McConnell}, D. 1997, \mnras, 289, 225

\bibitem[{{Taylor}(1999)}]{taylor99}
{Taylor}, A.~R. 1999, in ASP Conf. Ser. 168: New Perspectives on the
  Interstellar Medium, 3--89

\bibitem[{{Taylor} \& {Cordes}(1993)}]{taylor93}
{Taylor}, J.~H. \& {Cordes}, J.~M. 1993, \apj, 411, 674

\bibitem[{{Tenorio-Tagle} {et~al.}(1987){Tenorio-Tagle}, {Franco},
  {Bodenheimer}, \& {Rozyczka}}]{tenorio87}
{Tenorio-Tagle}, G., {Franco}, J., {Bodenheimer}, P., \& {Rozyczka}, M. 1987,
  \aap, 179, 219

\bibitem[{Thilker(1999)}]{thilkerphd}
Thilker, D.~A. 1999, PhD thesis, New Mexico State University

\bibitem[{{Thilker} {et~al.}(1998){Thilker}, {Braun}, \&
  {Walterbos}}]{thilker98}
{Thilker}, D.~A., {Braun}, R., \& {Walterbos}, R.~M. 1998, \aap, 332, 429

\bibitem[{{Tomisaka}(1998)}]{tomisaka98}
{Tomisaka}, K. 1998, \mnras, 298, 797

\bibitem[{{Toomre}(1963)}]{toomre63}
{Toomre}, A. 1963, \apj, 138, 385

\bibitem[{{Vall{\' e}e}(2002)}]{vallee02}
{Vall{\' e}e}, J.~P. 2002, \apj, 566, 261

\bibitem[{{Wada} \& {Norman}(1999)}]{wada99}
{Wada}, K. \& {Norman}, C.~A. 1999, \apjl, 516, L13

\bibitem[{{Wada} {et~al.}(2000){Wada}, {Spaans}, \& {Kim}}]{wada00}
{Wada}, K., {Spaans}, M., \& {Kim}, S. 2000, \apj, 540, 797

\bibitem[{Walter \& Brinks(1999)}]{walter99}
Walter, F. \& Brinks, E. 1999, \aj, 118, 273

\bibitem[{Walter {et~al.}(2001)Walter, Taylor, H\"uttemeister, Scoville, \&
  McIntyre}]{walter01}
Walter, F., Taylor, C.~L., H\"uttemeister, S., Scoville, N., \& McIntyre, V.
  2001, \aj, 121, 727

\bibitem[{{Weaver} {et~al.}(1977){Weaver}, {McCray}, {Castor}, Shapiro, \&
  {Moore}}]{weaver77}
{Weaver}, R., {McCray}, R., {Castor}, J., Shapiro, P., \& {Moore}, R. 1977,
  \apj, 218, 377

\end{thebibliography}
\normalsize

\clearpage

\clearpage

\clearpage
\begin{deluxetable}{lccccccl}
\tabletypesize{\scriptsize}
\tablewidth{0pt}
\tablecaption{Observed properties of \HI\ shells in the SGPS
\label{tab:shells}}
\tablehead{
\colhead{Name} &\colhead{$l$} & \colhead{$b$}& \colhead{$v$\tablenotemark{a}} & \colhead{$\Delta l$} & \colhead{$\Delta b$} & \colhead{$v_{\rm exp}$} & \colhead{Comments} \\
\colhead{} & \colhead{(deg)} & \colhead{(deg)} & \colhead{(${\rm km~s^{-1}}$)} & \colhead{(deg)} & \colhead{(deg)} & \colhead{(${\rm km~s^{-1}}$)} & \colhead{}
}
\startdata
GSH $255-00+52$ & $255.2$ & $-0.5$ & $+52$ & 3.8 & 3.7 & 18 &\\
GSH $256+00+63$ & $256.6$ & $-0.1$ & $+63$ & 2.2 & 2.2 & 12 &\\
GSH $263+00+47$ & $262.9$ & $-0.1$ & $+47$ & 2.3 & 2.9 & 13 & irregular shape \\
GSH $267-01+77$ & $267.1$ & $-1.1$ & $+77$ & 5.3 & 3.5 & 18 & two merged shells\\
GSH $270-03+42$ & $270.5$ & $-3.3$ & $+42$ & 3.2 & 3.8 & 12 &\\
GSH $277+00+36$\tablenotemark{b} & $277.5$ & $0.0$ & $+36$ & 5.4 & $>20$ & 20 & proposed chimney\\
GSH $280+00+59$\tablenotemark{b} & $279.8$ & $+0.1$ & $+59$ & 2.6 & $>20$ & 15 & 
proposed chimney\\
GSH $285-02+86$ & $285.6$ & $-2.5$ & $+86$ & 3.3 & 3.2 & 21 & \\
GSH $292-01+55$ & $292.4$ & $-1.5$ & $+55$ & 5.1 & 2.0 & 21 & \\
GSH $297-00+73$ & $297.2$ & $-0.3$ & $+73$ & 5.5 & 2.8 & 21 & two merged shells\\
GSH $298-01+35$ & $297.7$ & $-0.5$ & $+35$ & 0.8 & $>12$ & 10 & proposed chimney
 \\
GSH $304-00-12$\tablenotemark{c} & $303.9$ & $-0.2$ & $-12$ & 29 & 20 & 9 & Coalsack shell\\
GSH $305+01-24$\tablenotemark{c} & $305.1$ & $+1.0$ & $-24$ & 7 & $11.3$ & 6 & assoc. w/ Cen OB1\\
GSH $316-00+65$ & $316.4$ & $-0.5$ & $+65$ & 4.1 & 3.0 & 16 & two merged shells\\
GSH $319-01+13$ & $319.4$ & $-1.1$ & $+13$ & 2.1 & 1.7 & 18 &\\
GSH $327+04-25$ & $327.2$ & $+3.4$ & $-25$ & 7.7 & 5.6 & 7 &\\
GSH $337+00-05$ & $335.9$ & $+0.2$ & $-5$ & 18.6 & 12.5 & 9 & assoc. w/Ara OB1a\\
GSH $342-02-27$ & $341.8$ & $-1.9$ & $-27$ & 1.8 & 1.9 & 14 & \\
GSH $345+00+30$ & $345.2$ & $+0.5$ & $+30$ & 4.7 & 2.5 & 18 & unknown distance\\
\enddata
\tablenotetext{a}{All velocities all quoted with respect to the local standard of rest (LSR).}
\tablenotetext{b}{From \citet{mcgriff00}.}
\tablenotetext{c}{From \citet{mcgriff01c}.}
\end{deluxetable}

\clearpage
\begin{deluxetable}{lcccccccc}
\tabletypesize{\scriptsize}
\tablewidth{0pt}
\tablecaption{Derived properties for new \HI\ shells from the SGPS
\label{tab:dershells}}
\tablehead{
\colhead{Name} &\colhead{$D$} & \colhead{$R_{gal}$} & \colhead{$R_{\rm major}$} &\colhead{$R_{\rm minor}$} & \colhead{$t_D$} & \colhead{$n_0$} & \colhead{$M$} & \colhead{$E_{E}$} \\
\colhead{} & \colhead{(kpc)} & \colhead{(kpc)} & \colhead{(pc)} & \colhead{(pc)}& \colhead{(Myr)} &\colhead{($cm^{\rm -3}$)} & \colhead{($10^5~{\rm M_{\odot}}$)}& \colhead{($10^{51}$ ergs)}
}
\startdata
GSH $255-00+52$ & $5.3\pm1.0$ & $11.1\pm0.7$ & $175\pm 30$ & $170\pm30$ & 2.9& $1.7\pm0.5$ & $13\pm7$ & $53\pm39$ \\
GSH $256+00+63$ & $6.5\pm1.0$ & $11.9\pm0.8$ & $145\pm20$ & $140\pm20$ &3.6& $1.0\pm0.4$ & $4.2\pm1.9$ & $9\pm7$ \\
GSH $263+00+47$ & $5.5\pm1.0$ & $10.7\pm0.6$ & $110\pm20$ & $140\pm25$ &2.5& $1.4\pm0.6$ & $2.8\pm1.5$ & $7\pm5$\\
GSH $267-01+77$ & $9.2\pm1.2$ & $12.9\pm0.9$ & $420\pm55$ & $ 280\pm35$ &6.9& $0.4\pm0.1$ & $45\pm17$ & $177\pm104$ \\
GSH $270-03+42$ & $6.0\pm1.0$ & $10.4\pm0.6$ & $170\pm30$ & $ 200\pm30$ &4.2& $0.8\pm0.2$ & $5.3\pm2.6$ & $11\pm8$ \\
GSH $277+00+36$ & $6.5\pm0.9$ & $10.0\pm0.5$ & $305\pm45$ & $>1000$& 4.5& $1.2\pm0.1$ &27-56 & $240\pm120$ \\
GSH $280+00+59$ & $9.4\pm0.9$ & $11.5\pm0.7$ & $215\pm20$ & $>1000$ & 4.8& $0.6\pm0.1$ & $11\pm2$ & $26\pm 10$ \\
GSH $285-02+86$ & $13.7\pm1.3$ & $14.0\pm1.0$ & $395\pm40$ & $375\pm35$ &5.6 & $0.5\pm0.1$ & $44\pm13$ & $218\pm93$ \\
GSH $292-01+55$ & $11.6\pm1.0$ & $11.5\pm0.8$ & $515\pm45$ & $200\pm20$ &7.3& $0.5\pm0.1$ & $92\pm24$ & $464\pm196$ \\
GSH $297-00+73$ & $14.7\pm1.0$ & $13.3\pm1.0$ & $710\pm50$ & $360\pm25$ & 10.0& $0.3\pm0.1$ & $140\pm29$ & $190\pm67$ \\
GSH $298-01+35$ & $10.9\pm0.9$ & $10.2\pm0.7$ & $75\pm6$ & $>1200$ &2.2& $4.1\pm1.3$ & $2.3\pm0.6$ & $11\pm5$ \\
GSH $304-00-12$ & $1.2\pm0.6$\tablenotemark{a}& $7.9\pm0.4$ & $280\pm140$ & $200\pm100$  & 9.2& \nodata  & $19\pm8$ & $31\pm28$ \\ 
GSH $305+01-24$ & $2.2\pm0.6$\tablenotemark{a} &  $7.4\pm0.3$ & $140\pm40$ & $220\pm60$  & 6.9& \nodata  & $3.9\pm1.1$ & $8.3\pm7.1$ \\
GSH $316-00+65$ & $19.5\pm1.8$ & $14.5\pm1.5$ & $700\pm65$ & $500\pm45$ & 13.0&  $0.1\pm0.1$ & $69\pm19$ & $15\pm7$ \\
GSH $319-01+13$ & $14.0\pm0.9$ & $9.3\pm0.7$ & $ 260\pm15$ & $200\pm15$ & 6.4& $0.5\pm0.1$ & $13\pm3$ & $48\pm14$\\
GSH $327+04-25$ & $1.9\pm0.6$\tablenotemark{a} & $7.0\pm0.5$ & $130\pm40$ & $90\pm30$  & 5.6& \nodata & $2.2\pm0.7$ & $3\pm3$\\
GSH $337+00-05$ & $0.6\pm0.9$\tablenotemark{a} & $7.9\pm0.7$ & $90\pm150$ & $ 60\pm100$  & 3.0& \nodata & $1.0\pm1.7$ & $1.6\pm7$ \\
GSH $342-02-27$ & $2.6\pm0.7$\tablenotemark{b} & $6.1\pm0.6$ & $40\pm10$ & $40\pm10$  & 0.9& \nodata & $0.1\pm0.09$ & $0.25\pm0.23$ \\
GSH $345+00+30$ & $25.8\pm6.6$\tablenotemark{c} & $17.5\pm4.6$ & \nodata  & \nodata & \nodata & \nodata  & \nodata & \nodata \\
\enddata
\tablenotetext{a}{For dual-valued distances the distance determination, where
possible, is discussed in the text.}
\tablenotetext{b}{It was not possible to resolve the distance ambiguity for
this shell so parameters are calculated for both distances in the text.  In
the interest of space, only parameters calculated for the near distance are
given in the table.}
\tablenotetext{c}{Distance very uncertain}
\end{deluxetable}

\end{document}